\documentclass[letterpaper]{article} % DO NOT CHANGE THIS
\usepackage{aaai25}  % DO NOT CHANGE THIS
\usepackage{times}  % DO NOT CHANGE THIS
\usepackage{helvet}  % DO NOT CHANGE THIS
\usepackage{courier}  % DO NOT CHANGE THIS
\usepackage[hyphens]{url}  % DO NOT CHANGE THIS
\usepackage{graphicx} % DO NOT CHANGE THIS
\usepackage{natbib}  % DO NOT CHANGE THIS AND DO NOT ADD ANY OPTIONS TO IT
\usepackage{caption} % DO NOT CHANGE THIS AND DO NOT ADD ANY OPTIONS TO IT
\usepackage{algorithm}
\usepackage{algorithmic}

\usepackage{enumitem}
\usepackage{xspace}
\usepackage{thm-restate}
\usepackage{xcolor}         % colors
\usepackage{amsmath,amsfonts,amssymb,amsthm,mathrsfs,mathtools,dsfont}

\usepackage{newfloat}
\usepackage{listings}
\DeclareCaptionStyle{ruled}{labelfont=normalfont,labelsep=colon,strut=off} % DO NOT CHANGE THIS
\floatstyle{ruled}
\newfloat{listing}{tb}{lst}{}
\floatname{listing}{Listing}
\title{Robust Performance Incentivizing Algorithms for Multi-Armed Bandits with Strategic Agents}
\author {
    Seyed A. Esmaeili\textsuperscript{\rm 1},
    Suho Shin\textsuperscript{\rm 2},
    Aleksandrs Slivkins\textsuperscript{\rm 3}
}
\affiliations {
    \textsuperscript{\rm 1} University of Chicago\\
    \textsuperscript{\rm 2} University of Maryland\\
    \textsuperscript{\rm 3} Microsoft Research New York City\\
    esmaeili@uchicago.edu, suhoshin@umd.edu, slivkins@microsoft.com
}
\usepackage{bibentry}
\DeclareMathOperator*{\argmax}{arg\,max}

\DeclareMathOperator{\indic}{\mathds{1}}

\DeclareMathOperator{\E}{\mathbb{E}}

\newcommand{\floor}[1]{\left\lfloor #1 \right\rfloor}
\newcommand{\ceil}[1]{\left\lceil #1 \right\rceil}

\theoremstyle{plain}
\newtheorem{theorem}{Theorem}[section]

\newtheorem{lemma}[theorem]{Lemma}
\newtheorem{fact}[theorem]{Fact}

\newtheorem{claim}[theorem]{Claim}

\newtheorem{condition}[theorem]{Condition}

\newtheorem*{theorem*}{Theorem}
\newtheorem*{lemma*}{Lemma}
\newtheorem{claim*}{Claim}

\theoremstyle{definition}
\newtheorem*{definition*}{Definition}
\newtheorem{definition}[theorem]{Definition}

\theoremstyle{remark}

\newtheorem*{remark*}{Remark}
\newtheorem*{property*}{Property}

\newtheorem*{problem*}{Problem formulation}

\usepackage{stackengine}

\DeclareMathOperator{\bk}{\mathnormal{k_{\text{top}}}}     %\DeclareMathOperator{\bk}{\mathnormal{\bar{k}}}

\newcommand{\empi}[1]{\mathnormal{\hat{\mu}_i({#1})}}
\newcommand{\emp}[1]{\mathnormal{\hat{\mu}_{#1}}}

\DeclareMathOperator{\armt}{\mathnormal{I}_t}

\DeclareMathOperator{\arms}{\mathcal{A}}

\DeclareMathOperator{\armsmp}{\mathcal{A}_{\mathnormal{m\prime}}}

\DeclareMathOperator{\Dist}{\mathcal{D}}

\DeclareMathOperator{\meani}{\mathnormal{\mu_i}}

\newcommand{\totpulls}[1]{\mathnormal{T_{#1}}}
\newcommand{\toteffort}[1]{\mathnormal{C_{#1}}}
\newcommand{\hmeani}{\mathnormal{\hat{\mu}_i}}
\newcommand{\meanit}[1]{\mathnormal{\hat{\mu}_i(#1)}}

\newcommand{\ucbi}[1]{\mathnormal{UCB_{#1}}}

\newcommand{\estmean}[1]{\mathnormal{\hat{\mu}_{#1}}}

\DeclareMathOperator{\armsup}{\mathcal{A}_{\text{top}}}

\DeclareMathOperator{\armstop}{\mathcal{A}_{\text{top}}}

\DeclareMathOperator{\armsr}{\mathcal{A}_\mathnormal{r^*}}

\DeclareMathOperator{\maxi}{\mathnormal{M_i}}
\DeclareMathOperator{\maxall}{\mathnormal{M_{\text{top}}}}  %\DeclareMathOperator{\maxall}{\mathnormal{\bar{M}}}

\DeclareMathOperator{\spucb}{\mathnormal{S^{\text{SP+SAMF}}}}

\DeclareMathOperator{\spure}{\mathnormal{S^{\textsc{Pure-SP}}}}

\DeclareMathOperator{\pullseq}{\mathcal{P}}

\newcommand{\Sbesthet}{\mathnormal{S^*}} % {\mathnormal{\bar{S}^*_M}}
\newcommand{\Sbestheti}[1]{\mathnormal{S^*_{#1}}}

\newcommand{\Sbestheto}{\mathnormal{{S'}}}

\DeclareMathOperator{\ce}{\term{Clean \ Event}}
\DeclareMathOperator{\be}{\term{Bad \ Event}}

\newcommand{\mic}{\mathnormal{M_i^{f}}}

\newcommand{\kic}{\mathnormal{k_{top}}}
\DeclareMathOperator{\armstopc}{\mathnormal{\mathcal{A}_{\text{top}}}}
\DeclareMathOperator{\maxallc}{\mathnormal{M_{\text{top}}}}  %\DeclareMathOperator{\maxall}{\mathnormal{\bar{M}}}

\newcommand{\pc}[1]{\mathnormal{\text{PullCount}_{#1}}} 

\newcommand{\tildr}{\mathnormal{\tilde{r}}}

\DeclareMathOperator{\topt}{\mathcal{T}^{\text{OPT}}}
\DeclareMathOperator{\tsubopt}{\mathcal{T}^{\text{sub}}}

\newcommand{\toptj}[1]{\mathcal{T}^{\text{OPT}}_{#1}}
\newcommand{\tsuboptj}[1]{\mathcal{T}^{\text{sub}}_{#1}}

\newcommand{\poptj}[1]{\mathcal{P}^{\text{OPT}}_{#1}}
\newcommand{\psuboptj}[1]{\mathcal{P}^{\text{sub}}_{#1}}

\DeclareMathOperator{\honestarm}{\mathnormal{h}^*}

\DeclareMathOperator{\honestgapi}{\mathnormal{\honestmean - \ \mu_i}}

\DeclareMathOperator{\honestmean}{\mu^{*}_{\mathcal{H}}}

\DeclareMathOperator{\honestset}{\mathnormal{\arms}_{\mathcal{H}}}
\DeclareMathOperator{\armshonest}{\mathnormal{\arms}_{\mathcal{H}}}

\DeclareMathOperator{\barm}{\mathnormal{\bar{m}}}

\DeclareMathOperator{\Distup}{\mathcal{D}_{\mathnormal{r \ge \barm}}}
\DeclareMathOperator{\Distdown}{\mathcal{D}_{\mathnormal{r < \barm}}}

\DeclareMathOperator{\probdown}{\mathnormal{\Pr(r<\barm)}}
\DeclareMathOperator{\probup}{\mathnormal{\Pr(r \ge \barm)}}

\DeclareMathOperator{\meandown}{\mathnormal{\mu_{r <\barm}}}
\DeclareMathOperator{\meanup}{\mathnormal{\mu_{r \ge \barm}}}

\newcommand{\compInfo}{\text{public-information}\xspace}

\newcommand{\CompInfoInitCap}{\text{Public-Information}\xspace}

\newcommand{\compamb}{\text{private-information}\xspace}

\newcommand{\CompAmbInitCap}{\text{Private-Information}\xspace}

\newcommand{\term}[1]{\ensuremath{\mathtt{#1}}\xspace}

\DeclareMathOperator{\armseq}{\mathcal{A}_{\text{prev}}}
\DeclareMathOperator{\armsnon}{\mathcal{A}_{\text{non-prev}}}

\DeclareMathOperator{\Seq}{\mathnormal{S}_{\text{prev}}}
\DeclareMathOperator{\Snon}{\mathnormal{S}_{\text{non-prev}}}

\DeclareMathOperator{\Seqi}{\mathnormal{S}_{(\text{prev},\mathnormal{i})}}
\DeclareMathOperator{\Seqnoti}{\mathnormal{S}_{(\text{prev},\mathnormal{-i})}}

\begin{document}

\maketitle

\begin{abstract}
Motivated by applications such as online labor markets we consider a variant of the stochastic multi-armed bandit problem where we have a collection of arms representing strategic agents with different performance characteristics. The platform (principal) chooses an agent in each round to complete a task. Unlike the standard setting, when an arm is pulled it can modify its reward by absorbing it or improving it at the expense of a higher cost. The principle has to solve a mechanism design problem to incentivize the arms to give their best performance. However, since even with an effective mechanism agents may still deviate from rational behavior, the principal wants a robust algorithm that also gives a non-vacuous guarantee on the total accumulated rewards under non-equilibrium behavior. In this paper, we introduce a class of bandit algorithms that meet the two objectives of performance incentivization and robustness simultaneously. We do this by identifying a collection of intuitive properties that a bandit algorithm has to satisfy to achieve these objectives. Finally, we show that settings where the principal has no information about the arms' performance characteristics can be handled by combining ideas from second price auctions with our algorithms.
\end{abstract}

\section{Introduction}
Various settings of the web economy involve a platform (principal) who selects workers (agents) to complete tasks. These agents differ in their intrinsic performance abilities and can in addition exert further effort to improve their performance at the expense of a higher cost. Clearly, if the principal selects the top performing agents that give their best effort he would obtain higher revenue. To do that the principal has to incentivize higher quality performance through effective \emph{mechanism design/performance incentivization}. Furthermore, a deeper look would lead to a second important consideration, namely \emph{robustness}. While mechanism design can be effective in settings where the agents are entities like firms and corporations who exhibit high levels of rationality, when the agents are instead average individuals, deviations from rational behavior are known and observed to exist. Therefore, a more effective solution should not only incentivize higher quality contributions, but should also be robust; adaptively learning and adjusting to the agents' performance in order to generate a non-trivial amount of revenue even when the agents deviate from rational behavior.

We model this problem as a multi-armed bandit (MAB) problem where the agents (workers) are the arms. As in the stochastic MAB setting each arm has a reward (performance) distribution associated with it. However, we allow the agent to modify his rewards after the realization. More specifically, the agent is able to modify the reward by absorbing (stealing) the reward leading to a low quality output or he may improve (lift up) the reward to give a better output to the principal. Since real-life agents are heterogeneous in their skill level, in our model each agent has a specific mean for his distribution and can lift up his rewards to a specific (not necessarily unique) maximum value. To have a more realistic model of the agents, we assume that there always exists a subset of \emph{honest} agents, who either do not modify their rewards or possibly improve them.   

In this paper, we show how to simultaneously achieve both desiderata of performance incentivization and robustness. Specifically, we show that a class of MAB algorithms which we call \emph{sharply-adaptive, monotonic, and fair (SAMF)} are both performance incentivizing and robust. As the name suggests, SAMF algorithms satisfy a collection of properties which we show to be both natural and intuitive. A SAMF algorithm leads to an equilibrium where agents essentially give their top performance leading the principal to obtain optimal revenue. In fact, under a mild assumption on the size of market (number of agents) we show that no sub-optimal revenue equilibria can exist when using SAMF algorithms. We further consider a setting where the top performance level across the agents is not known either to the principal or the agents. We show that this setting can be handled by combining second price auction methods with SAMF algorithms, with the final mechanism leading to high revenue at equilibrium. Importantly, when agents deviate from equilibrium behavior, SAMF algorithms are robust and achieve revenue at least at the level of the honest agent with the highest reward mean. An advantage of our approach is its modularity. That is, we identify a set of properties that a MAB algorithm needs to satisfy to be a SAMF algorithm. Interestingly, we show that the standard UCB and $\epsilon$-greedy algorithms are examples of SAMF algorithms. In comparison to prior work, our setting allows for more complicated scenarios where agents may deviate from strategic behavior and interestingly we show that our algorithms can still achieve non-vacuous guarentees despite that. 
\section{Related Work}
There is an increasing attention in the multi-armed bandit literature to study the dynamics of algorithms in the presence of strategic behavior.
The most relevant paper, to the best of our knowledge, is due to \citet{feng2020intrinsic}, which considers a problem in which each arm can increase its reward given a budget constraint over the time horizon. Importantly however, the improvements the agents add to the reward in that setting are fake (strategic manipulations) although the principal does not realize that. The main objective of that paper is to see how susceptible MAB algorithms are to such manipulations. The main result of that paper is that standard algorithms like UCB, $\epsilon$-greedy, and Thompson sampling are intrinsically robust to those strategic manipulations given that sub-optimal arms are equipped with a sublinear amount of manipulation budget. This is different from our model since an increase in the reward in our setting brings a real improvement and the principal wants to induce good equilibria where essentially all agents perform at their top level. 
%Further, the bounds on the number of pulls an arm receives given a budget (although applicable to our setting) are too loose since we allow linear budgets.  

% if the principal would be fooled into pulling the.
% This is in stark contrast to our model as the principal in our setup in fact needs to motivate each arm to increase its reward by exerting more cost, since the principal enjoys the entire amount of delivered reward.
% The fake rewards in~\citet{feng2020intrinsic} in turn affects the algorithm to confuse as if those arms are equipped with truly good mean rewards.
% In this context, the principal aims to design an algorithm that is robust to this kind of adversarial behavior.
% They reveal that the standard algorithms like UCB, $\eps$-greedy, and Thompson sampling are intrinsically robust to those strategic manipulations given that each arm is equipped with a sublinear amount of manipulation budget.

\citet{braverman2019multiarmed} also study a problem of a similar flavor to ours. 
They consider each arm as a strategic agent who wants to maximize the total rewards she obtains through the horizon. Once an arm is pulled, the corresponding reward is realized and the agent can strategically take a certain fraction of the reward and deliver the remainder to the principal. Each agent here tries to maximize the cumulative reward by balancing the trade-off between (i) increasing short-term revenue by taking large fractions of the rewards and (ii) increasing the probability to be selected by taking a small fraction of the rewards.
They show that there are instances where any adversarial MAB algorithm would obtain a total of sublinear rewards (revenue). Essentially, the core cause behind this is that the arms can possibly engage in tacit collusion and therefore agree to take most of the rewards and send the principal a sublinear amount. In our setting, we assume a blind observation model (see Subsection \ref{subsec:model} for the definition) which makes the arms unable to collude to achieve such strategies since they cannot see the arm selections. Further, \citet{braverman2019multiarmed} construct an algorithm that recovers the second-highest mean of the arms based on a second-price auction algorithm. Our algorithm under a certain regime (of information) is inspired by their algorithm, but is more involved as we combine it with a MAB algorithm. This makes our algorithm (unlike theirs) robust, i.e., having a minimum (non-vacuous) revenue guarantee even when the arms follow non-equilibrium behaviour.  

There also exists a line of literature that considers strategic behavior in user-generated content platforms like Quora.
\citet{jain2009designing} study how to incentivize strategic users in user-generated content platforms to  contribute their own content immediately instead of postponing them.
\citet{ghosh2013learning}, preceded by \citet{ghosh2011game,ghosh2011incentivizing}, consider a problem of incentivizing high-quality contributions in user-generated content platforms, where each user wants to maximize her utility by exerting an optimized level of effort in constructing her content.
A major difference from our model is that in the previous papers the user's strategic choice is static rather than dynamic and made in a one-shot manner when she registers the content (mean of the arm).
% The registered contents are not modified after its submission, and each incoming agent is assumed to be myopic in the sense that they only interact with the platforms once, whereas the registered platforms might be kept pulled by the platform.
Another set of papers are focused on capturing different aspects of the strategic interaction in the multi-armed bandit problem - such as \citet{kremer2014implementing,mansour2015bayesian,bahar2020fiduciary} of incentivized exploration, \citet{shin2022multi,esmaeili2023replication} of strategic replication, \citet{wang2018multi} of known-compensation - however, we do not discuss the details as their models are significantly different from ours.

Our model is also conceptually related to the problem of incentivizing strategic workers to exert more costly effort also known as contract theory ~\citet{holmstrom1979moral}. 
% The standard model of contract theory follows from~\citet{holmstrom1979moral} which considers a principal-agent problem in which the principal faces an information asymmetry arising from moral hazards such as only observing the outcome of the agent's action but not the action itself.
% The agent's action induces a random outcome, and the principal commits to a contract which specifies a monetary reward for each outcome.
% Each action is equipped with a costly effort and the agent strategically chooses an action to maximize her own payoff.
% The principal needs to design a contract to maximize his expected utility assuming the agent's strategic behavior.
Designing an optimal and robust contract has been studied in various papers such as \citet{carroll2015robustness},~\citet{dutting2019simple}, and ~\citet{castiglioni2022designing}, and even in a repeated environment such as ~\citet{rogerson1985repeated} and \citet{chandrasekher2015unraveling}.
Our model might be interpreted as a repeated contract design with multiple agents. Our model, however, does not assume a random mapping between the agent's action and the outcome, and precludes monetary transfers but algorithmically incentivizes the agents to behave more in favor of the principal.

\section{Model, Objectives, and Main Results}\label{sec:model}
% \SE{1-emphasizes information} \\ 
% \SE{2-for now the number of arms $k$ is a constant} \\ 
% \SE{3-remove uneeded parts.}
% \begin{center}
% \begin{tabular}{ c c  }
%     Symbol & Definition  \\ 
%     $n$ & horizon  \\  
%     $k$ & \# of arms \\
%     $\totpulls{i}(t-1)$ & \# of times arm $i$ was pulled up and including round $t-1$ \\
%     $\toteffort{i}(t-1)$ & total amount of effort arm $i$ spent including in round $t-1$ \\
%     $\arms$ & set of arms \\
%     $M_i$ & maximum value arm $i$ can lift its reward to\\
%     $S_i$ & strategy followed by arm $i$ \\
%     $S=(S_1,\dots,S_k) $ & strategy profile followed by the arms \\
%     $S^*_i$ & best performance strategy for arm $i$  \\
% \end{tabular}
% \end{center}

% Here we start by giving a model of the interaction between the platform and the agents, then we formalize the objectives that the principal is concerned with. Finally, we give a summary of our results. 

\subsection{Model Details}\label{subsec:model}
We model the agents participating in a platform as a set of $k$ arms $\arms$\footnote{For simplicity, we assume that $k$ is a constant.}. The interaction runs over a collection of rounds from $1,2,\dots, n$ with $n$ being the \emph{horizon}. The principal (algorithm) would like to obtain top level performance from the arms while the arms would like to be pulled (selected) as many times as possible while spending as little effort as possible on improving their performance. 
% Now we give the full model details.

\paragraph{\textbf{Reward Modification:}} At round $t$, the algorithm chooses/pulls an arm (agent) $I_t$ and a reward $r_{I_t}(t) \in [0,1]$ is sampled from the static distribution $\Dist_{\armt}$ whose mean is $\mu_{\armt}$. After seeing the reward, arm $I_t$ can modify the reward by adding a value of effort $c_{I_t}(t)$ leading to the final reward $\Tilde{r}_{I_t}(t)=r_{I_t}(t)+c_{I_t}(t)$. Note that the algorithm only observes the reward  $\Tilde{r}_{I_t}(t)$. If $c_{I_t}(t) \ge 0$, then the reward is improved whereas if $c_{I_t}(t) < 0$ then the reward is degraded. Clearly, $c_{I_t}(t) < 0$ implies that arm $I_t$ has absorbed (a part of) its reward. Reward absorption is used to model the fact that in online labor markets a worker may intentionally worsen his performance\footnote{I.e., the worker may submit a low quality or even random output on the task to minimize the cost and time spent on it. Note that low quality output is often found in online labor markets, see for example \cite{ipeirotis2010quality,wais2010towards}. 
% \sscomment{but then, isn't the reward still the same? (if reward being the payment)}
}. Note that since we assume a setting where all realized and modified rewards are in $[0,1]$ then we always have $1-r_{I_t}(t) \ge c_{I_t}(t) \ge -r_{I_t}(t)$. We refer to the total effort spent by arm $i$ up to and including a given round $t$ by $\toteffort{i}(t)=\sum_{s=1}^t c_i(s)$. Given agent $i$ her strategy $S_i$ amounts to the values of $c_i(t)$ she chooses whenever she is pulled. Note that the agent's ability to modify the reward after seeing the realization gives the agent more control and follows related previous work such as \cite{feng2020intrinsic,braverman2019multiarmed}.  

\paragraph{\textbf{Utility:}} First, we denote the number of pulls arm $i$ receives up to and including round $t$ by $\totpulls{i}(t)$. Since the arm would like to be pulled more and spend as little cost as possible. A natural form for the utility is the following: 
\begin{align}\label{model:plain_util}
      u_i = \E[\totpulls{i}(n)] - \E[\toteffort{i}(n)]  
\end{align}

Clearly, from the previous paragraph on reward modification we see that reward absorption would lead to higher utility provided the arm still receives the same number of pulls while the reverse is true for reward improvement. This is in agreement with the fact that high quality effort is more costly. Note we may also generalize this by using a cost function leading to a utility of $u_i = \E[\totpulls{i}(n)] - \E[\sum_{t=1}^n f_i(c_i(t))] $ where $f_i(.)$ is a cost function specific to agent $i$. In the main body we mostly consider $f_i(c_i(t))=c_i(t)$. In Appendix \ref{sec:cost} we discuss how some results generalize to a wider family of cost functions. 

% Therefore the plain model assumes that the cost is equal to the effort spent. In \emph{cost (shape) function} utility model, then the cost would instead be $f_i(c_i(t))$ where $f_i(.)$ is a function that transforms the input effort into a final cost. Therefore the utility becomes:
% \begin{align}\label{model:cost_func_util}
%     \text{Cost Function Utility Model:} \ \ \ u_i = \E[\totpulls{i}(n)] - \E[\sum_{i=1}^n f_i(c_i(t))]  
% \end{align}

\paragraph{\textbf{Heterogeneous Characteristics of the Arms:}} It is natural to expect that the agents in the platform will have different levels of productivity which we model by allowing different characteristics for each agent (arm). Specifically, in addition to arms having different reward means, we assume that each arm $i$ can lift its reward to a maximum value of $M_i$. Further, $Support(\Dist_i) \subset [0,\maxi]$, this implies that $\mu_i \leq M_i$ and that if $i$ is pulled at round $t$ then $c_i(t) \leq M_i -r_i(t)$. In general, given two different arms  $i$ and $i'$ then we might have $M_i\neq M_{i'}$. We refer to the highest possible max reward value (\emph{top performance}) by $\maxall = \max_{i \in [k]} \maxi$. Further, the set of top performing arms is $\armstop = \{i\in [k]: \maxi = \maxall\}$ and their cardinality is $\bk=|\armstop|$. Moreover, we assume that the performance levels are not dependent on the horizon. I.e., $\forall i\in [k]: M_i =\Theta (1)$. 
%Further, we denote by $M_{\text{top-1}}$ the second highest performance level $M_i$.  
% Note that a special case is when all of the arms have the same $M_i$ value which we refer to as the homogeneous case.    

\paragraph{Honest Agents:} While it is true that in game-theoretic settings all agents are assumed to be rational utility maximizers \cite{roughgarden2010algorithmic,osborne2004introduction}, in reality we find that humans may deviate from utility maximizing strategies and possibly even be altruistic \cite{camerer1997progress,gintis2003explaining}. Therefore, in online labor markets which tend to include many participating agents it is natural to assume that a subset of the agents are \emph{honest}, meaning that they perform according to their productivity level and do not attempt to exploit the platform or give lower quality outputs. More formally, we assume that there always exists a set of \emph{honest agents} $\honestset \subset \arms$ with $|\honestset| \ge 1$ who always spend non-negative effort whenever they are pulled. I.e., if an honest agent $i$ is pulled at around $t$, then $c_i(t)\ge 0$ and therefore $\Tilde{r}_i(t) \ge r_i(t)$. Note that an honest arm can follow any strategy as long as the effort is non-negative ($c_i(t)\ge 0$). Further, we denote the maximum mean of reward distributions in the set of honest agents by $\honestmean$ and the honest agent with that maximum mean by $\honestarm$. I.e., $\honestmean= \max_{i \in \honestset} \mu_i$ and $\honestarm = \{i \in \honestset | \mu_i = \honestmean\}$. Moreover, in settings where we ask agents to ``report''\footnote{We will have settings where we elicit the agents to report their maximum value. However, in our setting reporting the value is done by pulling the arm and observing its reward which is different than for example auctions where a numeric value can be simply written by the agent.} their top performance value $M_i$, we assume that honest agents would always report the value truthfully.

Looking ahead, the guarantees of our algorithms still hold without honest agents under equilibrium settings obtaining essentially top revenue, but we cannot guarantee any robustness when agents deviate from equilibrium behavior. The latter are highly desirable due to the possibility of irrational behavior and mistakes. Further, we only make a mild assumption: having at least one honest agent, and will guarantee a strong ``robustness objective'' (see Sections \ref{sec:robustness_obj} and \ref{sec:robustness}) whereby our algorithms compete with the best honest agent in the wors case.

\paragraph{\textbf{Blind Observations:}} Unlike other settings such as advertisements, in online labor markets information about the decisions made by the platform are often unavailable or concealed. This is clearly the case in platforms such as Amazon Mechanical Turk or Uber where the agents do not observe the selections made by the platform. 
%More specifically, given a task in a platform such as Amazon Mechanical Turk or Uber an agent would not know when he is not chosen by the algorithm and when he is chosen he is unlikely to know his selection order. 
More formally, in a given round the agents who have not been selected (pulled) will not see which arm was pulled or the value of the reward it gave. Any arm is also not aware of the value of the round (temporal index) even when it is pulled. The arm is only aware of its own history (its own pulls, rewards, and effort). Although we consider this blind observation model, it remains non-trivial as the strategy of any arm is dynamic and dependent on its own history.  

\begin{figure}
    \centering
    \includegraphics[width=1.0\linewidth]{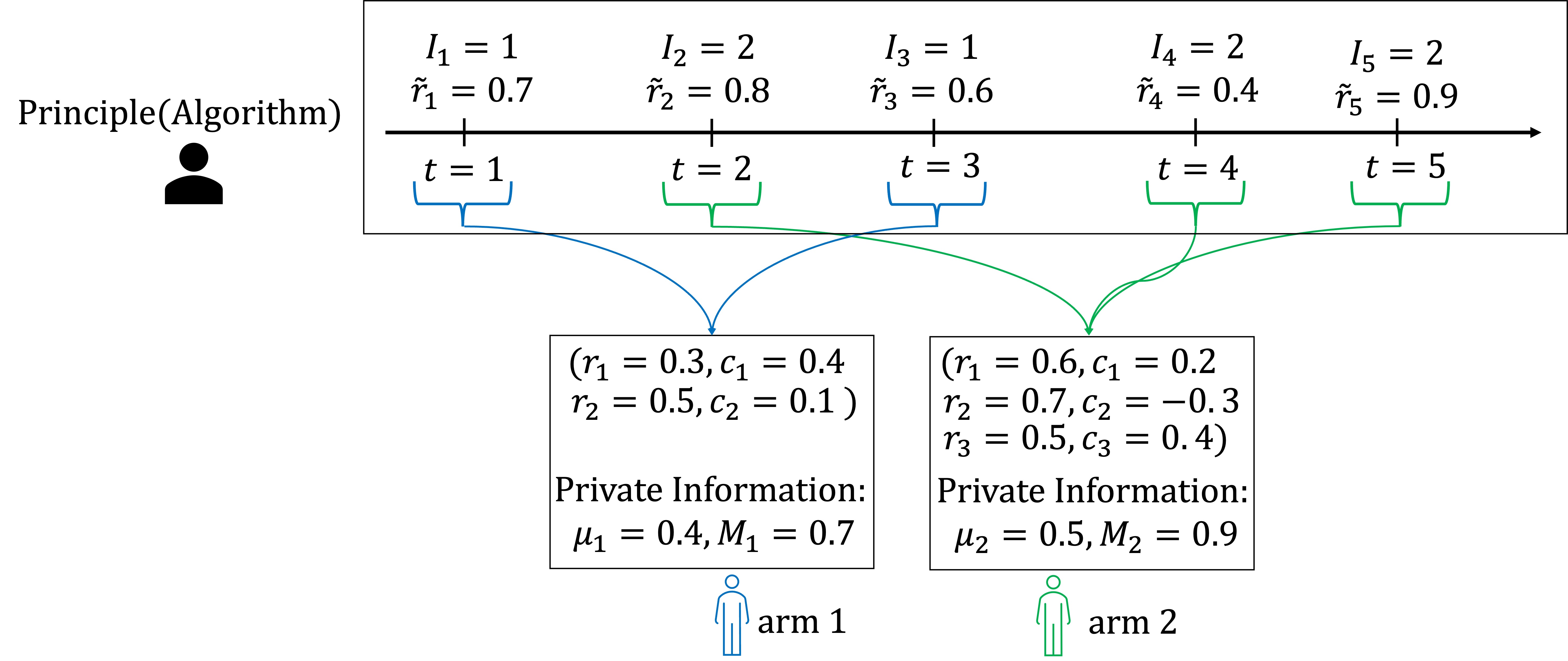}
    \caption{Model of the interaction. The information available to each agent and the principal (algorithm) is enclosed within their respective box. Since we are in a blind observation model, the agents do not have access to the same time index and use an ``internal'' time index that is different from that of the principle. Hence, agents 1 and 2 record their realized rewards and efforts by the indices $1,2$ and $1,2,3$, respectively.}
    \label{fig:model}
\end{figure}

Figure \ref{fig:model} shows an example of the interaction between the principle (algorithm) and agents in our model. Note how each agent has full knowledge of his own performance parameters ($\mu_i$ and $M_i$) unlike the principle. However, unlike the principle because of the blind observation model the agents lack information about the arm selections and the temporal index.

%Further, this model is practically motivated in settings such as online labor markets where the platform is not transparent and would not show which agent was selected and how much reward she gave.  

% \SE{remove this}
% Note that in a fully informed (actually even partially informed) observation model, the agents can engage in collusion and follow strategies that lead the principal to obtain very small rewards (i.e., $o(n)$ rewards) as shown in \cite{braverman2019multiarmed}. 
% \SE{end}

%\paragraph{$\CompInfoInitCap$ and $\CompAmbInitCap$ Settings:} 
\paragraph{Public and Private Information Settings:} 
As stated earlier the principal is unaware of any of the agents' performance characteristics (none of the $\meani$ and $M_i$ values). Further, while each agent $i \in \arms$ is aware of his own performance characteristics ($\meani$ and $M_i$), he is not aware of the other agents characteristics. This pauses a significant additional difficulty in mechanism design for performance incentivization similar to that encountered in auctions.   

We therefore consider two settings. The first, is the \emph{\textbf{\compInfo}} setting where the top performance level $\maxall$ is a \emph{public parameter known} by the principal and agents. Further, the market is large enough so that $\bk \ge 2$. Similar to auction theory, intuitively this setting alleviates the difficulty of the ``bidding'' issue among the top performing agents $\armstop$. However, designing an algorithm that is performance incentivizing while also being robust remains non-trivial. Moreover, this setting is strongly motivated by practical online labor markets such as Uber where it is known that a 5 star rating ($\maxall=5$) is achievable by many workers ($\bk\ge 2$). 

The second setting we consider is the \emph{\textbf{\compamb}} setting where the top performance level is not known neither to the agents nor the principal, thereby not requiring the strong assumptions of the public-information setting. To better represent our algorithms we start with the \compInfo setting in Section \ref{sec:compInfo} then the \compamb setting in Section \ref{sec:hetro_case_unique}. 

\subsection{Mechanism Design Objective and Equilibria}\label{subsec:objs}
The principal receives the rewards and therefore his primary measurement is the accumulated reward (revenue) through the horizon $P(n)$ defined as: 
\begin{align}
     P(n) = \E[\sum_{t=1}^n \Tilde{r}_{I_t}(t)]
\end{align}
Now we give the details for the principal's mechanism design objective. 

\paragraph{Performance Incentivization:} 
In light of the above model, the first natural objective for the principal is to accumulate revenue at the level of $\maxall$, i.e., have $P(n) \ge \maxall n- o(n)$. To do that the principal has to use an algorithm (mechanism) that is \emph{performance incentivizing}, i.e., obtaining top performance revenue at equilibrium. More specifically, let the strategy profile over the arms be $S=(S_1,\dots,S_k)$ then we should have: 
\begin{align}
     P(n,S) \ge \maxall n- o(n), \text{where $S$ is an equilibrium profile.} \nonumber 
\end{align}
We write $P(n,S)$ when we want to emphasize the revenue's dependence on the strategy. Now we formalize our notions of equilibrium. Note that our notions have similarities to that of \cite{braverman2019multiarmed}: 
\begin{definition}\label{def:one_approx_equilib}
\textbf{Asymptotic Equilibrium:} A strategy profile $S$ is an asymptotic equilibrium if given any arm $i \in [k]$ we have: 
\begin{align}
    \lim_{n \to \infty} \frac{u_i(S'_i, S_{-i})}{u_i(S_i, S_{-i})} \leq 1   \ \ \ \ \ \text{for any strategy $S'_i$.}
\end{align}
\end{definition}
The above essentially states, that at equilibrium an agent is not significantly better off as the horizon becomes larger by deviating to another strategy. 

Because of the heterogeneity in the agents' maximum performance levels, it is meaningful to introduce a more generalized notion of equilibrium. First, we establish some notation. Consider a partition of the set of arms $\arms$ into $\armseq$ (the ``prevalent'' set) and $\armsnon$ (the ``non-prevalent'' set) where the strategy profile is denoted by $\Seq$ and $\Snon$ for the sets $\armseq$ and $\armsnon$, respectively. Further, given an arm $i \in \armseq$, we denote the strategy profile the other arms in the prevalent set ($\armseq - \{i\}$) follow by $\Seqnoti$ whereas $i$'s strategy is $\Seqi$ under profile $\Seq$. We now define partial asymptotic equilibrium: 
\begin{definition}\label{def:partial_approx_equilib}
\textbf{Partial Asymptotic Equilibrium:} A partial asymptotic equilibrium holds if the set of arms $\arms$ can be partitioned into $\armseq \cup \armsnon$ such that $|\armseq| \ge 1$ with the arms in $\armseq$ following profile $\Seq$ such that the following is true $\forall i \in \armseq$: 
\begin{flalign}
    &  \lim_{n \to \infty} \frac{u_i(S'_i, \Seqnoti, S'_{\text{non-prev}})}{u_i(\Seqi, \Seqnoti,\Snon)} \leq 1 \\ 
    & \text{for any strategy $S'_i$ of $i$ and any strategy profile $S'_{\text{non-prev}}$} \nonumber  \\
    & \text{over $\armsnon$.} \nonumber  
\end{flalign}
\end{definition}
The notion of partial asymptotic equilibrium essentially states that the arms in the prevalent set are at equilibrium regardless of the strategies followed by the non-prevalent set. This notion captures a natural setting where the arms which cannot give top performance ($M_i < \maxall$) have little influence on the other arms and revenue.  

Note that for an arm $i$ the strategy $\Sbestheti{i}$ denotes the maximum performance strategy of always giving $M_i$. Further, $S^*$ denotes a strategy profile where either all of the arms $\arms$ or at least the subset of top arms $\armstop$ follow the strategy of giving the maximum reward value $\Sbestheti{i}$.    

\subsection{Robustness Objective}\label{sec:robustness_obj}
%\paragraph{Robustness}
While the above performance incentivization objective is clearly important and desirable, it is not sufficient. Specifically, it ignores non-equilibrium behavior which is known to be exhibited by agents in the real world. For example, agents may behave irrationally or they maybe non-strategic which is observed to happen in many settings \cite{camerer1997progress,guth1982experimental,singh2012investor}. Therefore, one would want the algorithm to be \emph{\textbf{robust}}, i.e., achieving as a fallback a non-trivial amount of revenue if non-equilibrium strategies are followed. Based on our model, a natural choice would be that the principal always obtains rewards at least at the level of highest mean of honest agents. Formally, we should have:
\begin{align}
    P(n,S) \ge \honestmean n- o(n) \ \ \text{for any strategy profile $S$.}
\end{align}

\begin{figure}
    \centering
    \includegraphics[width=0.7\linewidth]{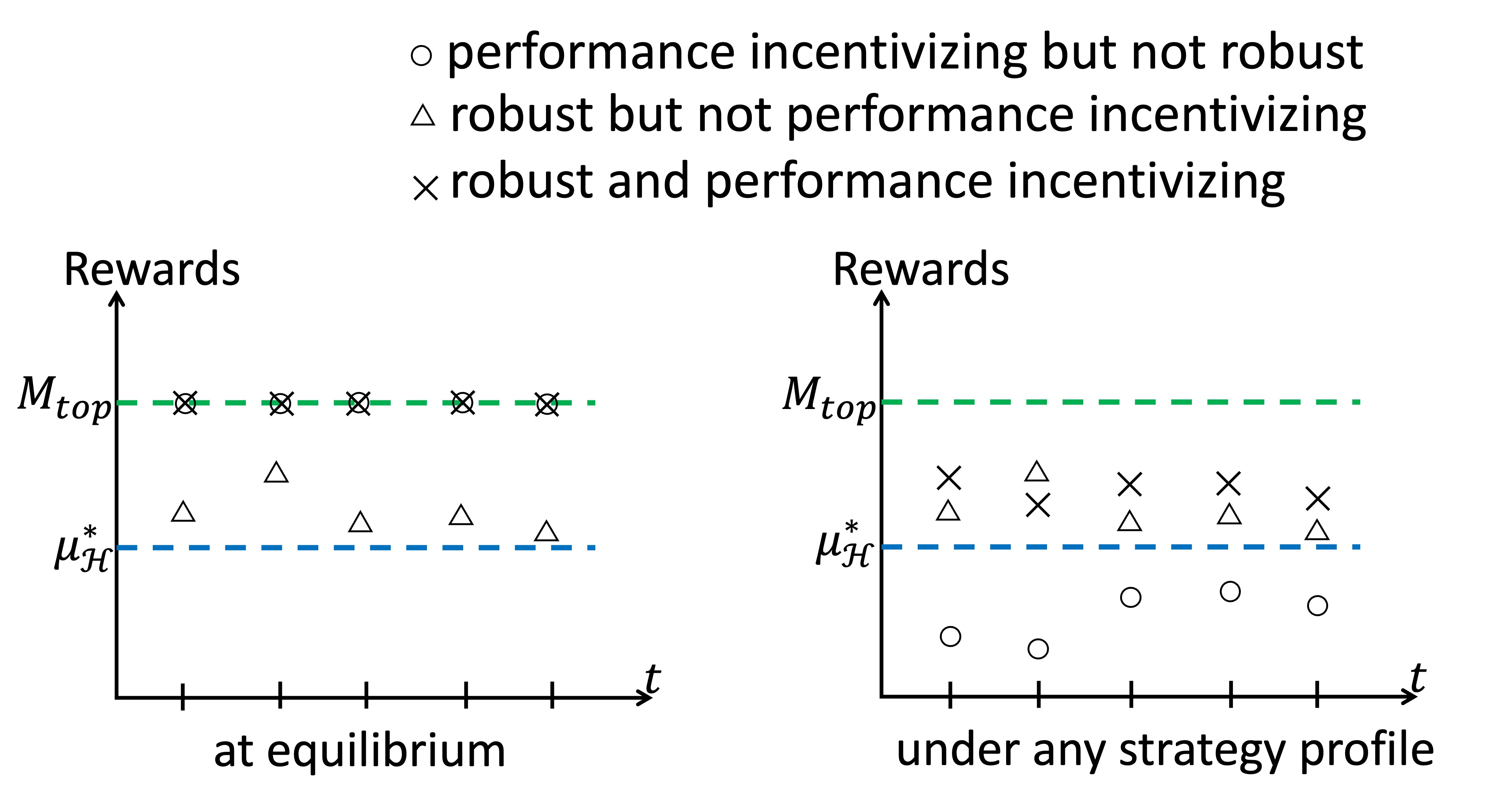}
    \caption{An illustrative figure showing the rewards we expect to obtain using different algorithms that satisfy different objectives. Notice how the algorithm that is both robust and performance incentivizing never falls below $\honestmean$ and obtains rewards of $\maxall$ at equilibrium.}
    \label{fig:objs}
\end{figure}

Figure \ref{fig:objs} shows an illustration of the two desired objectives of performance incentivization and robustness. 

\subsection{Main Results}\label{subsec:mainresults}
Our main results are the following:
\begin{itemize}[leftmargin=*]
    \item \textbf{Robustness:} In Section \ref{sec:robustness} we identify a property called \emph{sharp adaptivity} and show that any sharply adaptive MAB algorithm obtains revenue (total rewards) of at least $P(n) \ge \honestmean n-o(n)$ for any strategy profile. We further show that UCB and $\epsilon$-greedy are examples of sharply adaptive MAB algorithms.
    \item \textbf{Public-Information Setting:} In Section \ref{sec:compInfo} we identify two additional properties of MAB algorithms \emph{monotinicity} and \emph{fairness among the top arms (FATA)}, we abbreviate algorithms that satisfy these two properties in addition to sharp adaptivity by \emph{SAMF}. We prove that UCB and $\epsilon$-greedy are also SAMF algorithms. For the public-information setting we show that: (1) Under a condition which roughly corresponds to the number of top arms $\bk$ being large, no equilibrium that leads to revenue less than $\maxall n -o(n)$ exists when using SAMF algorithms. (2) SAMF algorithms lead to an equilibrium where all arms in $\armstop$ give the top performance regardless of whether the previous condition holds. 
    \item \textbf{Private-Information Setting:} In Section \ref{sec:hetro_case_unique} for the private-information setting we introduce an algorithm (SP+SAMF) that combines ideas from second price auctions with SAMF algorithms. This algorithm leads to an equilibrium with revenue of $M_{\text{top-1}}n-o(n)$ where $M_{\text{top-1}}$ is the second highest value of $M_i$.  \item \textbf{Failure of a Pure Mechanism Design Approach:} In Section \ref{sec:failure_of_pure_md} we underscore the importance of the robustness objective. Specifically, we construct a benign example where a non-robust mechanism obtains a vacuous amount of revenue despite being performance incentivizing. 
\end{itemize}
Further in Appendix \ref{sec:cost} we show how some of our results generalize in the public-information setting to a range of cost functions. Due to space limits we delay the proofs to the appendix.

\section{Obtaining a Robust Bandit Mechanism}\label{sec:robustness}
Here we show how to satisfy robustness. Specifically, we define a property which we call \emph{sharp adaptivity} and show that any sharply adaptive bandit algorithm is robust. I.e., it achieves revenue of $P(n)\ge \honestmean n-o(n)$ for any strategy profile. At an intuitive and rough level when using a sharply adaptive algorithm if an arm $i$ is pulled for a linear proportion of the horizon $\alpha n$ where $\alpha$ is a constant, then the total rewards accumulated must be at least at the level of $\honestmean$, if $i$ starts giving lower reward values then the algorithm quickly responds and no longer pulls $i$. Now we give the formal definition of sharp adaptivity. 
\begin{definition}\label{def:sr}
\textbf{Sharp Adaptivity}: A MAB algorithm is sharply adaptive if for any given arm $i\neq \honestarm$, if $\E[T_i(n)] \ge \alpha n$ where $\alpha>0$ is a constant, then the total expected effort spent by arm $i$ is at least $\E[\toteffort{i}(n)] \ge \alpha (\honestgapi) n - o(n)$, for any possible strategy profile followed by the arms. 
\end{definition}

% According to the above different sharply adaptive algorithms may differ in how adaptive they are. Specifically, if $i \neq \honestarm$ and  $\E[T_i(n)] \ge \alpha n$ then a sharply adaptive algorithm would have a cost lower bound of $\E[\toteffort{i}(n)] \ge \alpha (\honestgapi) n - h(n)$ where $h(n)=o(n)$. Clearly, a smaller $h(n)$ is preferred. When we want to characterize the quality of a sharply algorithm we say $h(n)$-sharply adaptive with $h(n)$ specified. 

% had two sharply adaptive algorithms then the first may have a cost lower bound of $\E[\toteffort{i}(n)] \ge \alpha (\honestgapi) n - h_1(n)$ whereas the other may have a bound of $\E[\toteffort{i}(n)] \ge \alpha (\honestgapi) n - h_2(n)$. Clearly, if $h_1(n) < h_2(n)$ then algorithm 1 is preferred. In general, we call an algorithm $h(n)$-sharply adaptive if 

We now prove that all sharply adaptive MAB algorithms satisfy robustness. This essentially follows since sharp adaptivity implies that if any arm $i \neq \honestarm$ is pulled for a linear proportion of the horizon ($\alpha n$ where $\alpha$ is some constant) then it must have given rewards of at least $\honestmean$ for that proportion ($\honestmean \alpha n -o(n)$), hence the total accumulated rewards through the horizon are at least $\honestmean n -o(n)$.  
\begin{restatable}{theorem}{theoremrobust}\label{th:sr_always_sb_regret}
For any strategy profile and arbitrary cost functions, a sharply adaptive MAB algorithm obtains revenue of $P(n) \ge \honestmean n -o(n)$.
\end{restatable}
% \begin{proofsketch}
% % A rough sketch of the proof is that when using a sharply adaptive algorithm pulling any arm for a linear proportion of the horizon ($\alpha n$ where $\alpha$ is some constant) must imply that the arm gave rewards of at least $\honestmean$ for that proportion ($\honestmean \alpha n -o(n)$). Therefore, the total accumulated rewards through the horizon are at least $\honestmean n -o(n)$.   
% \end{proofsketch}

% \SE{edit also note that we handle the fact that it is an r.v. not just a bound} Now we prove that UCB is sharply adaptive. We note that while \cite{feng2020intrinsic} proved a related robustness result for UCB, however the analysis is too loose. More concretely, in our setting their result implies that if $C_i(n) \leq \alpha (\honestgapi) n$, then $\E[T_i(n)|C_i(n)] \leq \beta \alpha  n$ where $\beta\ge 3$. Our results here on the other hand, are more refined satisfying sharp adaptivity instead of the more loose robustness notion of \cite{feng2020intrinsic}. Further, our results require a more careful analysis to show that if $\E[T_i(n)|C_i(n) \leq U]\ge \alpha n-o(n)$ where $U$ is some upper bound, then $U\ge \alpha (\honestgapi) n-o(n)$ with matching coefficients for the linear term. Without this refinement (matching linear coefficients), we cannot conclude that the algorithm would be robust.   

Interestingly, we prove that UCB and $\epsilon$-greedy are sharply adaptive. We note that while \cite{feng2020intrinsic} proved a related robustness result, their analysis is too loose. More concretely, in our setting their result implies that if $C_i(n) \leq \alpha (\honestgapi) n$, then $\E[T_i(n)] \leq \beta \alpha  n$ where $\beta\ge 3$. A bound with $\beta>1$ is not sufficient as it is too loose to guarantee robustness. We therefore have to establish the robustness of UCB and $\epsilon$-greedy using different techniques.

%A further advantage of our analysis is that it considers arbitrary possibly randomized strategies instead of only strategies with an upper bound on  $C_i(n)$. 

% Our results here on the other hand, are more refined satisfying sharp adaptivity instead of the more loose robustness notion of \cite{feng2020intrinsic}\footnote{The robustness notion of \cite{feng2020intrinsic} is essentially concerned with the order of an upper bound on $C_i(n)$ and $\E[T_i(n)]$, i.e. if $\E[T_i(n)]=\Omega(n)$ then $C_i(n)=\Omega(n)$ without being more exact about the coefficients in the linear order.}. Our refined analysis is necessary to establish our notion of robustness ($P(n)=\honestmean n-o(n)$ for any strategy profile). Further, unlike their setting we do not just assume an upper bound on $C_i(n)$ but treat it as a random variable and hence we lower bound $\E[C_i(n)]$. 

%Note that unlike UCB, $\epsilon$-greedy does not achieve the optimal regret bound in stochastic bandits\footnote{We use the version of $\epsilon$-greedy which does not assume knowledge of minimum gap mean gap ($\min_{i \in \arms} \mu_i^*-\mu_i$) as shown in block \ref{alg:eps_greedy}. Although a version of $\epsilon$-greedy that assumes knowledge of the minimum mean gap exists \cite{auer2002finite}, the assumption of knowing the smallest mean gap is highly unrealistic.}. 

\begin{restatable}{theorem}{ucbrobust}\label{th:UCB1_robust}
UCB and $\epsilon$-greedy are sharply-adaptive.  
\end{restatable}

Interestingly, one can also show that algorithms that are not sharply-adaptive can fail to be robust. Concrete examples include explore-then-commit and ``successive elimination'' (which is regret-optimal in the standard bandit setting) \cite{slivkins2019introduction}. This is because such algorithms only have an initial \emph{limited} phase (time period) where different arms are explored, and then afterwards an arm with the ``best performance'' is permanently selected and pulled for the rest of the horizon with the other arms being ``eliminated''. Therefore, if an arm with mean below $\honestmean$ gives best performance in that initial phase and then gives lower rewards (essentially below $\honestmean$) after being permanently selected, then it would clearly break robustness. 

%permanently select some arm and ``eliminate'' all others when this arm appears much better over some initial time period. 

\section{$\CompInfoInitCap$ Setting}\label{sec:compInfo}
% \begin{center}
% \begin{tabular}{ |c|c| } 
%  \hline
%  sharp adaptivity & Robust Mechanism  \\ 
%  sharp adaptivity + Monotonicity & Sub-optimal Profit Equilibria do not exist under condition \ref{cond:cond_good_revenue}  \\ 
%  sharp adaptivity + Monotonicity + FATA  & Optimal Profit Under Equilibrium regardless of condition  \ref{cond:cond_good_revenue} \\ 
%  \hline
% \end{tabular}
% \end{center}

% We start this section by showing an interesting impossibility result. Specifically, 
% no arm $i$ can have the maximum performance strategy $\Sbestheti{i}$ of always giving $M_i$ as a \emph{dominant strategy} if the principal uses a sublinear MAB algorithm. 
Having resolved the robustness objective by using a sharply-adaptive algorithm we now turn to the mechanism design issue in the public-information setting. We start this section by showing an interesting impossibility result. Specifically, for any sublinear regret MAB algorithm\footnote{As in the standard terminology, a sublinear regret MAB algorithm is one where in the ordinary stochastic setting (arms cannot modify their realized rewards) the regret is sublinear, i.e., $R(n)=(\max\limits_{i \in \arms} \mu_i) n - \E[\sum_{t=1} r_{I_t}(t)]=o(n)$.} that the principle can use no arm $i$ can have the maximum performance strategy $\Sbestheti{i}$ of always giving $M_i$ as a \emph{dominant strategy}. The intuition behind this result is that while an agent may receive most of the pulls in the horizon by giving the top performance level of $\maxall$, because of the performance cost he would obtain higher utility by lowering his performance to only outperform the second highest agent. 
\begin{restatable}{theorem}{domstra}\label{th:domstra}
Under any sublinear regret MAB algorithm, it is not a dominant strategy for an arm $i$ to follow the maximum performance strategy $S^*_i$. 
\end{restatable}

%%%%%%%% AUG 11, 2024
% Now, we provide our main results on the competition-informed setting. Similar to how we achieved robustness, we identify two additional properties \emph{monotonicity} and \emph{fairness among the top arms (FATA)} that are sufficient to induce a good equilibrium. We call algorithms with these properties \emph{sharply adaptive monotonic and fair (SAMF)} algorithms and show that standard algorithms like UCB and $\epsilon$-greedy are examples of SAMF algorithms. 

% We begin by introducing the monotonicity property. 

Before we provide our main positive results, we introduce two properties of bandit algorithms that lead to good equilibria. We start with monotonicity which essentially states that an arm $i$ is rewarded with more pulls if it upgrades its performance to the top performance level $S^*_i$.

% We first identify a set of properties for an algorithm to induce good equilibrium and show that standard algorithms, specifically UCB and $\epsilon$-greedy satisfy them.
% Then, under a mild assumption on the number of top performance arms, we show that there exists no equilibrium such that principal receives revenue less than $\maxall$ if one uses performance incentivizing algorithm.
% We complement this result by proving that without any condition on the number of top performance arms, any performance incentivizing algorithm admits a partial asymptotic equilibrium such that the principal obtains $\maxall n - o(n)$.

\begin{definition}\label{def:monotonic}
\textbf{Monotonicity}: A MAB algorithm satisfies monotonicity if for any strategy profile $S=(S_1,\dots,S_k)$, if an arm $i$ was to deviate to the top performance strategy $S^*_i$, then for any reward seed and any strategy seed\footnote{Note that a strategy is in general randomized.} $\forall j \neq i: \totpulls{j}(n,S^*_i,S_{-i}) \leq \totpulls{j}(n,S_i,S_{-i})$.
\end{definition}

Furthermore, we introduce the Fairness Among the Top Arms (FATA) property which states that arms that follow the top performance strategy roughly receive an equal number of pulls. 
%We now define the FATA property. 
\begin{definition}\label{def:fata}
\textbf{Fairness Among the Top Arms (FATA)}: When there exists a subset of arms $\armsr$ such that whenever $i \in \armsr$ is pulled then a reward of $r^*=\max_{t\in [n]} r_t$ is given, then a MAB algorithm is fair among the top arms (FATA) if $\forall i,i' \in \armsr: |\E[\totpulls{i}(n)]-\E[\totpulls{i'}(n)]|\leq o(n)$. 
\end{definition}
% The FATA property states that if there were arms that always give the maximum reward through the horizon whenever pulled, then they will receive almost the same number of pulls by the algorithm.  

Bandit algorithms that satisfy sharp adaptivity, monotonicity, and FATA will play a fundamental and we abbreviate them by \emph{SAMF}. 
% \sscomment{note: incentivizing and -sing are both used. Make it consistent}
\begin{definition}\label{def:pi}
\textbf{SAMF algorithm}: A MAB algorithm is SAMF if it
satisfies sharp adaptivity, monotonicity, and FATA.
\end{definition}

We now show that standard algorithms such as UCB and $\epsilon$-greedy are SAMF.

\begin{restatable}{theorem}{ucbrpi}\label{th:ucbrpi}
UCB and $\epsilon$-greedy are SAMF algorithms. 
\end{restatable}

In the next two subsections, we consider the $\compInfo$ setting and prove positive equilibrium results with SAMF algorithms. 
%The first result (Theorem \ref{th:sr_np_bad_equilib}) excludes the possibility of equilibria that are not performance incentivizing (leading to less than $\maxall n-o(n)$ revenue). The result holds if the number of top arms $\bk$ is sufficiently large. The second result (Theorem \ref{th:pi_equilib}) complements the first by showing that there always exists an equilibrium leading to top performance revenue regardless of the number of top arms $\bk$. 
%Note that since SAMF algorithms are sharply-adaptive it immediately follows that achieve robustness by Theorem \ref{th:sr_always_sb_regret}

% (Theorem \ref{th:sr_np_bad_equilib}) shows the non-existence of  This result holds under a mild assumption which essentially  
% \SE{here}
% Our first equilibrium result (Theorem \ref{th:sr_np_bad_equilib}) shows that when the number of top performance arms $\bk$ is sufficiently large, then there does not exist an equilibrium such that principal receives revenue less than $\maxall$ if one uses performance incentivizing algorithm. We complement this result by proving that without any condition on the number of top performance arms, any performance incentivizing algorithm admits a partial asymptotic equilibrium such that the principal obtains $\maxall n - o(n)$ (Theorem \ref{th:pi_equilib}).

\subsection{Non-Existence of Sub-Optimal Revenue Equilibria for Large $\bk$}\label{subsec:no_bad_equilb}
In the $\compInfo$ setting, we show that there exists no asymptotic equilibrium for which the principal receives revenue (total rewards) less than $\maxall$. Specifically, we show that any MAB algorithm that is SAMF\footnote{More specifically, this result holds if the MAB algorithm is sharply adaptive and monotonic but the algorithm does not necessarily need to satisfy the FATA property.} has no asymptotic equilibrium where a principal obtains a revenue less than $P(n)=\maxall n-o(n)$. Note that we actually show that no partial asymptotic equilibrium exists and therefore this implies that no asymptotic equilibrium exists where the principal obtains less than top performance revenue. All that is required is an assumption on the number of top arms as given below: 

\begin{condition}\label{cond:cond_good_revenue}
$\bk > \frac{2}{\min_{i \in \armstop} (1+\meani-\maxall)} (1+\epsilon)$ where $\epsilon>0$ is a constant. 
\end{condition}

%\paragraph{Interpretation of Condition \ref{cond:cond_good_revenue} and Theorem \ref{th:sr_np_bad_equilib}:} 

We now present the theorem which excludes bad equilibria. 
\begin{restatable}{theorem}{nobadequlib}\label{th:sr_np_bad_equilib}
Given a MAB algorithm that satisfies sharp adaptivity and monotonicity, if Condition \ref{cond:cond_good_revenue} is satisfied, then there exists no partial asymptotic equilibrium strategy profile $S$ such that $P(n,S)\leq \alpha \maxall n +o(n)$ where $\alpha <1$. \end{restatable}
%\begin{proofsketchandinterp}
\paragraph{Proof Sketch and Interpretation of Condition \ref{cond:cond_good_revenue}.}
At a rough level, the proof shows that there must be a linear proportion of the horizon of size at least $(1-\alpha) n-o(n)$ where sub-optimal revenue (less than $\maxall$) is accumulated. Further, there must be a top arm that is pulled for at most $\frac{1}{\bk}$ of that proportion (at most $\frac{(1-\alpha) n-o(n)}{\bk}$). If the number of top arms $\bk$ is large then the arm's utility is low. If the cost of performing at the top value which is $(\maxall-\meani)$ in expectation is relatively small compared to the current utility as would be implied by Condition \ref{cond:cond_good_revenue} then we would have $1-(\maxall-\meani) > \frac{(1+\meani)}{\bk}$, then this arm would have the incentive to deviate to a top performance level to take the $(1-\alpha) n-o(n)$ portion of the horizon. 

This theorem also implies that the principal can make sure that he always obtains top performance revenue by increasing the number of top arms (higher $\bk$) and by making the gap needed for optimal performance $(\maxall-\meani)$ small. 

Although the above explanation is intuitive, since the MAB setting is dynamic the proof of the theorem is involved and requires some new techniques. 

\subsection{Achieving Top Performance Equilibrium for Any Value of $\bk$}
Here we show that a SAMF algorithm leads to top performance revenue equilibrium even if Condition \ref{cond:cond_good_revenue} does not hold. Specifically, we show that $S^*$ is an equilibrium profile. At an intuitive level if a top arm $i \in \armstop$ follows strategy $\Sbestheti{i}$ of always giving $\maxall$ then by sharp adaptivity any other top arm $j \in \armstop-\{i\}$ needs to perform at that level (always give rewards of $\maxall$) to receive a linear number of pulls. By additionally invoking the the monotonicity and FATA properties we can establish upper and lower bounds on $j$'s utility through bounds on the number of pulls and cost. With the utility of a top arm $j$ lower bounded under $S^*$ and upper bounded under any deviation strategy the equilibrium is established.

\begin{restatable}{theorem}{compinfequilib}\label{th:pi_equilib}
Given the $\compInfo$ setting, if a MAB algorithm is SAMF, then the strategy profile $S^*$ where $\forall i \in \armsup: S_i=S^*_i$ is a partial asymptotic equilibrium leading to optimal revenue $P(n,S^*)=\maxall n -o(n)$. 
\end{restatable}

\section{$\CompAmbInitCap$ Setting: SAMF Algorithms with a Second Price Auction}\label{sec:hetro_case_unique}
%%%%% AUG 11th 
% Here we deal with the problem assuming that the principal and agents do not know the top performance value $\maxall$ not even the top arms $\armstop$. We are not in the $\compInfo$ but rather the $\compamb$ setting (Subsection \ref{subsec:model}). We show that this setting can be handled using an algorithm (mechanism) which combines elements from second price auctions with a SAMF algorithm. We assume that the SAMF algorithm we use has in the ordinary stochastic setting an instance-dependent logarithmic upper bound for the number of pulls of sub-optimal arms. I.e., in the ordinary stochastic (unmodifiable) reward setting, a sub-optimal arm $i$ has  $\E[T_i(n)] = O(\frac{\ln(n)}{\Delta^2_i})$. UCB is an example of such an algorithm. Our algorithm below tunes the parameters according to a logarithmic upper bound on the number of pulls, however other bounds can be accommodated as long as the MAB algorithm is a SAMF algorithm. 

Here we generalize our algorithm (mechanism) to deal with the case where the top performance level $\maxall$ is not publicly known neither to the principle nor to any of the arms\footnote{Note that as usual each arm $i$ still knows its properties including its own maximum performance level $M_i$. But in this setting it would not know if $\maxall=M_i$.}. Since our mechanism combines methods from second price (SP) auctions with SAMF algorithms we call it \textsc{SP+SAMF} (Algorithm \ref{alg:spa_UCB_mech}). The rounds under this mechanism can be divided into 3 phases: (\texttt{A}) bidding phase rounds (line 1), (\texttt{B}) SAMF phase (line 4), and (\texttt{C}) reward phase (line 5). The rounds in the first and last phases (\texttt{A}) and (\texttt{C}) are only $o(n)$ while the middle phase (\texttt{B}) contains $\Omega(n)$ rounds.  We assume that the SAMF algorithm we use has in the ordinary stochastic setting an instance-dependent logarithmic upper bound for the number of pulls of sub-optimal arms. I.e., in the ordinary stochastic (unmodifiable) reward setting, a sub-optimal arm $i$ has  $\E[T_i(n)] = O(\frac{\ln(n)}{\Delta^2_i})$. UCB is an example of such an algorithm. The mechanism tunes some parameters according to a logarithmic upper bound on the number of pulls, however other bounds can be accommodated as long as the MAB algorithm is a SAMF algorithm. 

% We show that this setting can be handled using an algorithm which combines elements from second price auctions with a SAMF algorithm.

% % %%%%%%%%%% HERE AAAI 2025
% Here we generalize our algorithm (mechanism) to deal with the case where the top performance level $\maxall$ is not publicly known neither to the principle nor to any of the arms\footnote{Note that as usual each arm $i$ still knows its properties including its own maximum performance level $M_i$. But in this setting it would not know if $\maxall=M_i$.}. We show that this setting can be handled using an algorithm which combines elements from second price auctions with a SAMF algorithm. We assume that the SAMF algorithm we use has in the ordinary stochastic setting an instance-dependent logarithmic upper bound for the number of pulls of sub-optimal arms. I.e., in the ordinary stochastic (unmodifiable) reward setting, a sub-optimal arm $i$ has  $\E[T_i(n)] = O(\frac{\ln(n)}{\Delta^2_i})$. UCB is an example of such an algorithm. Our algorithm below tunes the parameters according to a logarithmic upper bound on the number of pulls, however other bounds can be accommodated as long as the MAB algorithm is a SAMF algorithm. 
% % %%%%%%%%%% HERE AAAI 2025

%We note that in this private-information setting unlike the public-information setting we may have $\bk=1$. 

We give a brief intuitive explanation for the key ideas of the mechanism. The bidding process of phase (\texttt{A}) where each arm $i$ reports its top performance value by giving a reward $m_i$ resolves the difficulty of not knowing the top performance level. This is the case since the mechanism tells all of the arms the value of $m'$ which is the second highest value among the $m_i$ values\footnote{Note that if more than one arm reports the same maximum value then $m'$ would actually equal the maximum value.}. In phase (\texttt{B}), if $\bk \ge 2$ then the setting essentially reduces to the public-information setting where the strategy profile $S^*$ of always giving the top performance is an equilibrium. On the other hand, if $\bk =1$ then we show that it is an equilibrium for the unique top arm to give a performance value of $m'+\frac{1}{\ln(n)}$ which is slightly above the second highest value and for the other arms to give their top performance value of $M_i$. For the unique top arm, this equilibrium is established since outperforming the rest of the arms enables it to essentially receive almost of the pulls ($n-o(n)$ pulls) whereas for the other arms giving their top performance results in higher pulls in the last reward phase (\texttt{C}) since each arm is pulled a number of times proportional to its performance (average reward it gives) in phase (\texttt{B}). Critically, our mechanism include the blocking condition of line (4A) in phase (\texttt{B}). The main objective behind this blocking condition is to make sure that the top arms do not gain a higher utility by bidding untruthfully with a higher value (see Appendix \ref{app:blocking_cond} for a concrete example). Note that this blocking condition is ``punishing'' for untruthful bidding instead of under performance as done in \cite{braverman2019multiarmed}. In the next section we consider a mechanism based on \cite{braverman2019multiarmed} that punishes for under-performance instead. We demonstrate its failure due to its lack of robustness even under truthful bidding. 
%We show that while such a mechanism would be performance incentivizing it obtains only a sublinear amount of revenue with high probaility since it would not be robust. 
Since we assume that honest agents always bid truthfully and we use a SAMF algorithm in phase (\texttt{B}) we always achieve robustness for any strategy profile.

\begin{algorithm}[h!] %https://tex.stackexchange.com/questions/56871/how-to-format-for-loop-for-printing-a-pseudo-code-listing 
\caption{\textsc{SP+SAMF}}\label{alg:spa_UCB_mech}
    \begin{algorithmic}
    \STATE (1) Pull each arm $i$ once, let $m_i$ be its reported value. %{\textbackslash\textbackslash Bidding Phase }
    \STATE (2) Let $m'$ be the second highest reported value.
    \STATE (3) Tell all arms the value $m'$. \nonumber 
    \STATE (4) Ignore all previous rewards and use a SAMF MAB algorithm with two modifications: %{\textbackslash\textbackslash SAMF Phase}
    \STATE   \ \ \ \ \ \ 4A-if an arm $i$ has $m_i \leq m'$, but gives reward $r > m'$, then block the arm (never play it again). \\ 
    \STATE   \ \ \ \ \ \ 4B-stop at round $t=n- k \ceil{\ln(n)}^{\rho+3}$ for some fixed constant $\rho >0$.\\ 
    \STATE (5) Tell the arms that this is the reward phase then pull each arm $i$ for a number of rounds $N_i$ where $\E[N_i]=\hmeani^{\text{SAMF}} \big(\ln(n)\big)^{\rho+3}$ and $\hmeani^{\text{SAMF}}$ is the mean of arm $i$ during SAMF phase rounds. 
\end{algorithmic}
\end{algorithm}

% %%%%%%%%%% HERE AAAI 2025
% Before we present our equilibrium strategy, we note that the rounds can be divided into 3 phases:  bidding phase rounds (line 1), SAMF phase (line 4), and reward SAMF phase (line 5). Note that $k \ceil{\ln(n))}^{\rho+3}=o(n)$. Therefore, only the SAMF phase has a linear number of rounds $\Omega(n)$. Further, as noted the performance levels are not dependent on the horizon. I.e., $\forall i\in [k]: M_i =\Theta (1)$. Moreover, we denote by $M_{\text{top-1}}$ the second highest maximum performance level of $M_i$.  
% %%%%%%%%%% HERE AAAI 2025

% \SE{(1) In the strategy profile below, we have $m'+\frac{1}{\sqrt{n}}$. The value $\frac{1}{\sqrt{n}}$ is not actually best. The last payment rounds $k \ceil{\ln(n)}^4$ are also dependent on it. I tried optimizing it but the expression was tedious. For now, I thought it's worthwhile to show an algorithm with a good equilibrium that works and optimize it later. We are also assuming that the difference between the two highest performance levels $M_1$ and $M_2$ is a constant, hence the top arm can give $m'+\frac{1}{\sqrt{\ln(n)}}$. If that is not the case, then the arms could be essentially at the same level. (2) Some arms will have sublinear utility. Accordingly, the notion of equilibrium that should be used is $\lim_{n \to \infty} \frac{u_i(S'_i, S_{-i}|\mech)}{u_i(S_i, S_{-i}|\mech)} \leq 1$ as discussed in definitions (\ref{def:add_approx_equilib},\ref{def:one_approx_equilib}).}

% %%%%%%%%%% HERE AAAI 2025
Below we give a full description of the equilibrium strategy profile. For convenience, we assume below that if $\bk=1$, then the only agent with $M_i=\maxall$ is not $\honestarm$ (the honest agent with the maximum mean), this case leads to identical guarantees but using a more complicated strategy for the honest agent $\honestarm$ since we assume that honest agents would never absorb (degrade) their rewards (see Appendix \ref{app:unique_top_is_honest}).
% %%%%%%%%%% HERE AAAI 2025

\begin{restatable}{theorem}{thspucb}\label{thm:sp_ucb}
Under SP+SAMF (Algorithm \ref{alg:spa_UCB_mech}), the strategy profile $\spucb$ is an asymptotic equilibrium that consists of the following: (1) In the bidding phase (line 1) each arm bids truthfully with $m_i=M_i$, (2) In the reward phase (line 5), all rewards are absorbed if the agent is not honest otherwise no positive effort is added. (3) In the SAMF phase (line 4) the strategy is: 
\begin{align*}
  S_i^{\textsc{SP+SAMF}} =
  \begin{cases}
   \text{if $M_i \leq m'$, then always give a reward of $M_i$} \\ 
   \text{if $M_i > m'$, then give $m'+\frac{1}{\ln(n)}$ }\\
  \end{cases}
\end{align*}\end{restatable}

Denoting the second highest top performance value by $M_{\text{top-1}}$, the following theorem is immediately implied by Theorem~\ref{thm:sp_ucb}:
\begin{restatable}{theorem}{sprpiguaren}\label{th:sprpi}
The SP+SAMF mechanism (Algorithm (\ref{alg:spa_UCB_mech})) leads to: (1) an equilibrium where the principal obtains revenue of $P(n)\ge M_{\text{top-1}} \cdot n-o(n)$. (2) Revenue of $P(n)\ge \honestmean n-o(n)$ for any strategy profile.   
\end{restatable}

\section{Failure of a Non-Robust Mechanism Design Approach}\label{sec:failure_of_pure_md}
Here we show the failure of a non-robust mechanism design approach even in a setting where all agents are honest, we consider a mechanism similar to the one introduced in \cite{braverman2019multiarmed}. Specifically, consider \textsc{Pure-SP} (Algorithm \ref{alg:pure_sp_mech}) which uses elements from second price auction similar to \textsc{SP+SAMF} but unlike \textsc{SP+SAMF} does not use a SAMF algorithm. More concretely, the mechanism starts with a bidding phase and then tells all arms the second highest reported value $m'$. It then proceeds to pull all of the arms which gave the maximum bid value ($\max_{i \in \arms} m_i$) in a cycling manner. If a top bidding arm under-performs, it is deleted and not pulled again as shown in line (5A). The mechanism ends with a rewarding phase where each arm $i$ is pulled a number of times proportional to its bid value $m_i$.

\begin{algorithm}[h!] %https://tex.stackexchange.com/questions/56871/how-to-format-for-loop-for-printing-a-pseudo-code-listing 
\caption{\textsc{Pure-SP}}\label{alg:pure_sp_mech}
    \begin{algorithmic}
    \STATE (1) Pull each arm $i$ once, let $m_i$ be its reported value.. %{\textbackslash\textbackslash Bidding Phase }
    \STATE (2) Let $m'$ be the second highest reported value.
    \STATE (3) Tell all arms the value $m'$. \nonumber 
    \STATE (4) Let $\armsmp$ be the subset of arms that gave the highest $m_i$ value. I.e., $\armsmp=\{i \in \arms | m_i = \argmax_{j \in \arms} m_j\}$ 
    \STATE (5) Pull arms $i \in \armsmp$ in a cycling manner: 
           \STATE   \ \ \ \ \ \ 5A-delete arm $i$ if it gives a value less than $m'$ and update $\armsmp$.   
           \STATE   \ \ \ \ \ \ 5B-stop at round $t=n- 2k$. 
    \STATE (6) Tell the arms that this is the reward phase then pull each arm $i$ for $N_i$ rounds where $\E[N_i]=2m_i$.
\end{algorithmic}
\end{algorithm}
 
It is not difficult to show that this mechanism leads to a revenue of $P(n) \ge M_{\text{top-1}} n-o(n)$ under an equilibrium strategy profile (see Appendix \ref{app:pure_sp}). However, if the agents deviate from the equilibrium strategy the revenue can easily vanish. Specifically, in the theorem below we show that even in an instance where all agents are honest the revenue can be sublinear with high probability if the agents deviate from the equilibrium in a given pull with a small constant probability of $\epsilon$.

\begin{restatable}{theorem}{thpuresp}\label{thm:pure_sp}
For \textsc{Pure-SP} (Algorithm \ref{alg:pure_sp_mech}) there exists an instance where even if the bidding is done truthfully and all of the agents are honest, if an agent at a subsequent pull deviates from the equilibrium strategy with probability $\epsilon>0$ then with probability at least $1-\frac{1}{n}$ the revenue $P(n)=o(n)$. 
\end{restatable}

\bibliography{refs}

\begin{thebibliography}{28}
\providecommand{\natexlab}[1]{#1}

\bibitem[{Bahar et~al.(2020)Bahar, Ben-Porat, Leyton-Brown, and
  Tennenholtz}]{bahar2020fiduciary}
Bahar, G.; Ben-Porat, O.; Leyton-Brown, K.; and Tennenholtz, M. 2020.
\newblock Fiduciary bandits.
\newblock In \emph{International Conference on Machine Learning}, 518--527.
  PMLR.

\bibitem[{Braverman et~al.(2019)Braverman, Mao, Schneider, and
  Weinberg}]{braverman2019multiarmed}
Braverman, M.; Mao, J.; Schneider, J.; and Weinberg, S.~M. 2019.
\newblock Multi-armed Bandit Problems with Strategic Arms.
\newblock In Beygelzimer, A.; and Hsu, D., eds., \emph{Proceedings of the
  Thirty-Second Conference on Learning Theory}, volume~99 of \emph{Proceedings
  of Machine Learning Research}, 383--416. PMLR.

\bibitem[{Camerer(1997)}]{camerer1997progress}
Camerer, C.~F. 1997.
\newblock Progress in behavioral game theory.
\newblock \emph{Journal of economic perspectives}, 11(4): 167--188.

\bibitem[{Carroll(2015)}]{carroll2015robustness}
Carroll, G. 2015.
\newblock Robustness and linear contracts.
\newblock \emph{American Economic Review}, 105(2): 536--63.

\bibitem[{Castiglioni, Marchesi, and Gatti(2022)}]{castiglioni2022designing}
Castiglioni, M.; Marchesi, A.; and Gatti, N. 2022.
\newblock Designing Menus of Contracts Efficiently: The Power of Randomization.
\newblock \emph{arXiv preprint arXiv:2202.10966}.

\bibitem[{Chandrasekher(2015)}]{chandrasekher2015unraveling}
Chandrasekher, M. 2015.
\newblock Unraveling in a repeated moral hazard model with multiple agents.
\newblock \emph{Theoretical Economics}, 10(1): 11--49.

\bibitem[{D{\"u}tting, Roughgarden, and Talgam-Cohen(2019)}]{dutting2019simple}
D{\"u}tting, P.; Roughgarden, T.; and Talgam-Cohen, I. 2019.
\newblock Simple versus optimal contracts.
\newblock In \emph{Proceedings of the 2019 ACM Conference on Economics and
  Computation}, 369--387.

\bibitem[{Esmaeili, Hajiaghayi, and Shin(2023)}]{esmaeili2023replication}
Esmaeili, S.; Hajiaghayi, M.; and Shin, S. 2023.
\newblock Replication-proof Bandit Mechanism Design.
\newblock \emph{arXiv preprint arXiv:2312.16896}.

\bibitem[{Feng, Parkes, and Xu(2020)}]{feng2020intrinsic}
Feng, Z.; Parkes, D.; and Xu, H. 2020.
\newblock The intrinsic robustness of stochastic bandits to strategic
  manipulation.
\newblock In \emph{International Conference on Machine Learning}, 3092--3101.
  PMLR.

\bibitem[{Ghosh and Hummel(2011)}]{ghosh2011game}
Ghosh, A.; and Hummel, P. 2011.
\newblock A game-theoretic analysis of rank-order mechanisms for user-generated
  content.
\newblock In \emph{Proceedings of the 12th ACM conference on Electronic
  commerce}, 189--198.

\bibitem[{Ghosh and Hummel(2013)}]{ghosh2013learning}
Ghosh, A.; and Hummel, P. 2013.
\newblock Learning and incentives in user-generated content: Multi-armed
  bandits with endogenous arms.
\newblock In \emph{Proceedings of the 4th conference on Innovations in
  Theoretical Computer Science}, 233--246.

\bibitem[{Ghosh and McAfee(2011)}]{ghosh2011incentivizing}
Ghosh, A.; and McAfee, P. 2011.
\newblock Incentivizing high-quality user-generated content.
\newblock In \emph{Proceedings of the 20th international conference on World
  wide web}, 137--146.

\bibitem[{Gintis et~al.(2003)Gintis, Bowles, Boyd, and
  Fehr}]{gintis2003explaining}
Gintis, H.; Bowles, S.; Boyd, R.; and Fehr, E. 2003.
\newblock Explaining altruistic behavior in humans.
\newblock \emph{Evolution and human Behavior}, 24(3): 153--172.

\bibitem[{G{\"u}th, Schmittberger, and Schwarze(1982)}]{guth1982experimental}
G{\"u}th, W.; Schmittberger, R.; and Schwarze, B. 1982.
\newblock An experimental analysis of ultimatum bargaining.
\newblock \emph{Journal of economic behavior \& organization}, 3(4): 367--388.

\bibitem[{Holmstr{\"o}m(1979)}]{holmstrom1979moral}
Holmstr{\"o}m, B. 1979.
\newblock Moral hazard and observability.
\newblock \emph{The Bell journal of economics}, 74--91.

\bibitem[{Ipeirotis, Provost, and Wang(2010)}]{ipeirotis2010quality}
Ipeirotis, P.~G.; Provost, F.; and Wang, J. 2010.
\newblock Quality management on amazon mechanical turk.
\newblock In \emph{Proceedings of the ACM SIGKDD workshop on human
  computation}, 64--67.

\bibitem[{Jain, Chen, and Parkes(2009)}]{jain2009designing}
Jain, S.; Chen, Y.; and Parkes, D.~C. 2009.
\newblock Designing incentives for online question and answer forums.
\newblock In \emph{Proceedings of the 10th ACM conference on Electronic
  commerce}, 129--138.

\bibitem[{Kremer, Mansour, and Perry(2014)}]{kremer2014implementing}
Kremer, I.; Mansour, Y.; and Perry, M. 2014.
\newblock Implementing the “wisdom of the crowd”.
\newblock \emph{Journal of Political Economy}, 122(5): 988--1012.

\bibitem[{Mansour, Slivkins, and Syrgkanis(2015)}]{mansour2015bayesian}
Mansour, Y.; Slivkins, A.; and Syrgkanis, V. 2015.
\newblock Bayesian incentive-compatible bandit exploration.
\newblock In \emph{Proceedings of the Sixteenth ACM Conference on Economics and
  Computation}, 565--582.

\bibitem[{Mitzenmacher and Upfal(2017)}]{mitzenmacher2017probability}
Mitzenmacher, M.; and Upfal, E. 2017.
\newblock \emph{Probability and computing: Randomization and probabilistic
  techniques in algorithms and data analysis}.
\newblock Cambridge university press.

\bibitem[{Osborne et~al.(2004)}]{osborne2004introduction}
Osborne, M.~J.; et~al. 2004.
\newblock \emph{An introduction to game theory}, volume~3.
\newblock Oxford university press New York.

\bibitem[{Rogerson(1985)}]{rogerson1985repeated}
Rogerson, W.~P. 1985.
\newblock Repeated moral hazard.
\newblock \emph{Econometrica: Journal of the Econometric Society}, 69--76.

\bibitem[{Roughgarden(2010)}]{roughgarden2010algorithmic}
Roughgarden, T. 2010.
\newblock Algorithmic game theory.
\newblock \emph{Communications of the ACM}, 53(7): 78--86.

\bibitem[{Shin, Lee, and Ok(2022)}]{shin2022multi}
Shin, S.; Lee, S.; and Ok, J. 2022.
\newblock Multi-armed Bandit Algorithm against Strategic Replication.
\newblock In \emph{International Conference on Artificial Intelligence and
  Statistics}, 403--431. PMLR.

\bibitem[{Singh(2012)}]{singh2012investor}
Singh, S. 2012.
\newblock Investor irrationality and self-defeating behavior: Insights from
  behavioral finance.
\newblock \emph{Journal of Global Business Management}, 8(1): 116.

\bibitem[{Slivkins et~al.(2019)}]{slivkins2019introduction}
Slivkins, A.; et~al. 2019.
\newblock Introduction to multi-armed bandits.
\newblock \emph{Foundations and Trends{\textregistered} in Machine Learning},
  12(1-2): 1--286.

\bibitem[{Wais et~al.(2010)Wais, Lingamneni, Cook, Fennell, Goldenberg,
  Lubarov, Marin, and Simons}]{wais2010towards}
Wais, P.; Lingamneni, S.; Cook, D.; Fennell, J.; Goldenberg, B.; Lubarov, D.;
  Marin, D.; and Simons, H. 2010.
\newblock Towards building a high-quality workforce with mechanical turk.
\newblock \emph{Proceedings of computational social science and the wisdom of
  crowds (NIPS)}, 1--5.

\bibitem[{Wang and Huang(2018)}]{wang2018multi}
Wang, S.; and Huang, L. 2018.
\newblock Multi-armed bandits with compensation.
\newblock \emph{Advances in Neural Information Processing Systems}, 31.

\end{thebibliography}

\newpage 
\clearpage
\newpage 
\onecolumn
\appendix
\section{Full Pseudo Code for UCB and $\epsilon$-Greedy}

\begin{algorithm}[h!] %https://tex.stackexchange.com/questions/56871/how-to-format-for-loop-for-printing-a-pseudo-code-listing 
\caption{\textsc{UCB}}\label{alg:ucb}
    \begin{algorithmic}[1]
    \STATE Play each arm once for the first $k$ rounds.
    \STATE In a given round $t>k$:
    \STATE \ \ Pull arm $i$ with highest $UCB_i(t-1)=\hat{\mu}_i(t-1)+\sqrt{\frac{2 \ln(n)}{\totpulls{i}(t-1)}}$ (breaking ties arbitrarily) 
\end{algorithmic}
\end{algorithm}

\begin{algorithm}[h!] %https://tex.stackexchange.com/questions/56871/how-to-format-for-loop-for-printing-a-pseudo-code-listing 
\caption{\textsc{$\epsilon$-Greedy}}\label{alg:eps_greedy}
    \begin{algorithmic}[1]
    \STATE Play each arm once for the first $k$ rounds.
    \STATE Set $\epsilon=32 (k \ln(n))^{\frac{1}{3}} n^{\frac{-1}{3}}$. 
    \STATE In a given round $t>k$:  
    \STATE \ \ Sample a Bernoulli random variable $s_t$ from $Bern(\epsilon)$. 
    \STATE \ \ If $s_t=1$, then a pick an arm uniformly at random.
    \STATE \ \ If $s_t=0$, then sample an arm uniformly at random from the set of arms with maximum empirical mean.
\end{algorithmic}
\end{algorithm}
\section{Omitted Proofs}
We restate the first theorem and then give its proof: 
\theoremrobust* 
\begin{proof}
Consider any strategy profile $S$ then we have: 
\begin{align*}
    P(n) & = \sum_{i \in \arms} \sum_{t=1}^n \E[\Tilde{r}_i(t)] \\  
    & \ge \sum_{i \in \arms: \E[T_i(n)]=\Omega(n)} \ \sum_{t=1}^n \E[\Tilde{r}_i(t)] \\ 
    & = \sum_{i \in \arms: \E[T_i(n)]=\Omega(n)} \ \Big(\mu_i \E[T_i(n)] + \E[\sum_{t=1}^n c_i(t)] \Big) \\ 
    & \ge \sum_{i \in \arms: \E[T_i(n)]=\Omega(n)} \ \Big(\mu_i \E[T_i(n)] + (\honestgapi) \E[T_i(n)]- o(n) \Big) && (\text{by Definition \ref{def:sr} of sharp adaptivity}) \\ 
    & \ge \sum_{i \in \arms: \E[T_i(n)]=\Omega(n)} \ \Big( \honestmean \E[T_i(n)] - o(n) \Big) \\ 
    & \ge  \honestmean n - o(n)
\end{align*}
Note that since in the above we only used the accumulated rewards without invoking the cost functions, the proof holds even for arbitrary cost functions $f_i(.)$. 
% For stochastic bandits the regret can be written as $R(n)= \sum_{i} \di \E[\totpulls{i}(n)]$. But since effort is enabled in our model, the regret is $R(n) = \sum_{i} \Big(\di \E[\totpulls{i}(n)] - \E[C_i(n)] \Big)$. Clearly, to have linear regret $R(n) = \Omega(n)$, then a sub-optimal arm $i$ has to be pulled for a linear number of times. Therefore, $\E[\totpulls{i}(n)] = \alpha n + o(n)$, but definition (\ref{def:sr}) states that this can only happen if $\toteffort{i}(n)\ge \alpha \di n -o(n)$. Accordingly, arm $i$'s contribution to the regret would be $\di \E[\totpulls{i}(n)] - \E[C_i(n)] \leq \di ( \alpha n + o(n) ) - ( \alpha \di n -o(n) ) = o(n)$. 
\end{proof}

Now we prove that UCB is sharply adaptive: 
\begin{theorem}\label{th:ucbsa}
UCB is sharply adaptive.
\end{theorem}
\begin{proof}
Define the $\ce$ to be an event such that $\forall i \in \arms, \forall t \in [n]: |\meanit{t}-\mu_i| \leq \sqrt{\frac{2\ln(n)}{T_i(t)}}$. Let the $\be$ be the complement of the $\ce$.
We start by showing that the clean event holds with high probability: 
\begin{claim}\label{cl:ucb_clean_event}
  \begin{align*}
    \Pr(\ce) \ge 1 - \frac{2}{n^2} 
\end{align*}
\end{claim}
\begin{proof}
Given the infinite reward tapes, let $v_i(\ell)$ be the average reward of arm $i$ for $\ell$ pulls.
Then, by Hoeffding's inequality \ref{th:hoffbound} we have
\begin{align*}
    \Pr\Big(|v_i(\ell)-\mu_i| \ge \sqrt{\frac{2\ln(n)}{\ell}}\Big) \leq 2 \exp(-2 \ell \frac{2 \ln(n)}{\ell}) = \frac{2}{n^4}
\end{align*}
Using the union bound, we further obtain
\begin{align*}
    \Pr(\ce) \ge 1 - \frac{2 k n}{n^4} \ge 1-\frac{2}{n^2},
\end{align*}
and it completes the proof.
\end{proof}

Now we show that if arm $i$ has $\E[T_i(n)] \ge \alpha n$ where the constant $\alpha>0$ then the following claim holds: 
\begin{claim}\label{cl:pulls_ce_lb}
\begin{align*}
    \E[T_i(n)|\ce] \ge \alpha n - 1 
\end{align*}
\end{claim}
\begin{proof}
By the law of total expectation, observe that
\begin{align*}
        \E[T_i(n)] & = \E[T_i(n)|\ce] \cdot \Pr(\ce) + \E[T_i(n)|\be] \cdot \Pr(\be) \\ 
                   & \leq \E[T_i(n)|\ce] \cdot 1 + \frac{2}{n^2} \cdot n \tag{due to Claim~\ref{cl:ucb_clean_event}}\\
                   & = \E[T_i(n)|\ce]  + \frac{2}{n}.
\end{align*}
Hence, we obtain $\E[T_i(n)|\ce] \ge \E[T_i(n)] - \frac{2}{n} \ge \alpha n -1$ since $n \ge 2$. 
\end{proof}
Recall that under $\ce$, we have $\forall i \in \arms, \forall t \in [n]: |\meanit{t}-\mu_i| \leq \sqrt{\frac{2\ln(n)}{T_i(t)}}$. 
Then, it follows that for any $t \in [n]$, 
\begin{align}
    \ucbi{i}(t) &\leq \mu_i + 2 \sqrt{\frac{2\ln(n)}{T_i(t)}} + \frac{\sum_{s=1}^t c_i(s)}{T_i(t)} \label{eq:ucbi} \\ 
    \ucbi{\honestarm}(t) &\ge \honestmean \label{eq:ucbh}
\end{align}
% \begin{align}
%     \forall t \in [n]: \nonumber \\ 
%     & \ucbi{i}(t) \leq \mu_i + 2 \sqrt{\frac{2\ln(n)}{T_i(t)}} + \frac{\sum_{s=1}^t c_i(s)}{T_i(t)} \label{eq:ucbi} \\ 
%     & \ucbi{\honestarm}(t) \ge \honestmean \label{eq:ucbh}
% \end{align}
Consider a realization of the reward tapes such that $t_0$ is the index of last round that arm $i$ is pulled.
Since arm $i$ is pulled at round $t_0$, we have
\begin{align*}
    \ucbi{i}(t_0-1) & \ge \ucbi{\honestarm}(t_0-1).
\end{align*}
Using inequalities \eqref{eq:ucbi} and \eqref{eq:ucbh}, we obtain
\begin{align*}
    \mu_i + 2 \sqrt{\frac{2\ln(n)}{T_i(t_0-1)}} + \frac{\sum_{s=1}^{t_0-1} c_i(s)}{T_i(t_0-1)} & \ge \honestmean
\end{align*}
Rearranging the inequality, we have
\begin{align*}
    % \mu_i + 2 \sqrt{\frac{2\ln(n)}{T_i(t_0-1)}} + \frac{\sum_{s=1}^{t_0-1} c_i(s)}{T_i(t_0-1)} & \ge \honestmean  && \tag{(By inequalities \ref{eq:ucbi} and \ref{eq:ucbh})} \\ 
    % \iff 
    \sum_{s=1}^{t_0-1} c_i(s) \ge (\honestgapi) T_i(t_0-1) - 2\sqrt{2\ln(n) T_i(t_0-1)} 
\end{align*}
It follows that under this realization, $C_i(n) \ge \sum_{s=1}^{t_0-1} c_i(s) - 1$ since $c_i(t_0) \ge -1$. 
Note also that $T_i(n)=T_i(t_0-1)+1$ due to the construction of $t_0$.
This yields that
\begin{align*}
    C_i(n)  &  \ge (\honestgapi) T_i(n) - 2\sqrt{2\ln(n) T_i(n)} - (1+(\honestgapi))
\end{align*}
By taking the expectation over the $\ce$ we have
\begin{align*}
    \E[C_i(n)|\ce]  & \ge (\honestgapi) \E[T_i(n)|\ce] - 2 \E[\sqrt{2\ln(n) T_i(n)}|\ce] - (1+(\honestgapi)) \\
                     & \ge (\honestgapi) \E[T_i(n)|\ce] - 2 \sqrt{2\ln(n) \E[T_i(n)|\ce]} - (1+(\honestgapi)) \\
                     & \ge (\honestgapi) (\alpha n-1) - 2 \sqrt{2\ln(n) (\alpha n-1)} - (1+(\honestgapi)) \ \ \tag{by Claim \ref{cl:pulls_ce_lb}}
\end{align*}
% \begin{align*}
%     C_i(n)  &  \ge (\honestgapi) T_i(n) - 2\sqrt{2\ln(n) T_i(n)} - (1+(\honestgapi)) \\ 
%     & \text{By taking the expectation over the $\ce$ we have:} \\
%      \E[C_i(n)|\ce]  & \ge (\honestgapi) \E[T_i(n)|\ce] - 2 \E[\sqrt{2\ln(n) T_i(n)}|\ce] - (1+(\honestgapi)) \\
%                      & \ge (\honestgapi) \E[T_i(n)|\ce] - 2 \sqrt{2\ln(n) \E[T_i(n)|\ce]} - (1+(\honestgapi)) \\
%                      & \ge (\honestgapi) (\alpha n-1) - 2 \sqrt{2\ln(n) (\alpha n-1)} - (1+(\honestgapi)) \ \ \text{(by Claim \ref{cl:pulls_ce_lb})}
% \end{align*}
Note that in the second inequality, we use the fact that $f(x) = \sqrt{x}$ is a concave function and Jensen's inequality to derive 
\begin{align*}
    \E[\sqrt{2\ln(n) T_i(n)}|\ce]\leq \sqrt{2\ln(n) \E[T_i(n)|\ce]}.
\end{align*}
Putting all claims together, we finally obtain the following lower bound on $\E[C_i(n)]$:
\begin{align*}
    \E[C_i(n)] & = \E[C_i(n)|\ce] \cdot \Pr(\ce) + \E[C_i(n)|\be] \cdot \Pr(\be) \\ 
               & \ge \E[C_i(n)|\ce] \cdot \Big(1-\frac{2}{n^2}\Big) +  (-n) \cdot \frac{2}{n^2} \\ 
               & \ge (\honestgapi) (\alpha n-1) - 2 \sqrt{2\ln(n) (\alpha n-1)} - o(1) \\ 
               & \ge (\honestgapi) \alpha n - o(n).
\end{align*}
This finishes the proof.
\end{proof}

Now we prove that $\epsilon$-greedy is sharply adaptive: 
\begin{theorem}\label{th:epsgreedysa}
$\epsilon$-greedy is sharply adaptive.
\end{theorem}
\begin{proof}
We start by defining the $\ce$. Under the $\ce$ we have:
\begin{align}
\forall i \in \arms, \forall t \in [n]: \ \ & |\meanit{t}-\mu_i| \leq \sqrt{\frac{2\ln(n)}{T_i(t)}} \label{eq:ce_eps_1} \\ 
\forall i \in \arms, \forall t \ge  n^{\frac{2}{3}} k^{\frac{1}{3}} (\ln(n))^{\frac{1}{3}}: \ \ & T^R_i(t) \ge  \frac{c}{2} k^{-\frac{2}{3}} (\ln(n))^{\frac{1}{3}} n^{\frac{-1}{3}} t  \label{eq:ce_eps_2}
\end{align}
\end{proof}
We want to show that the clean event happens with high probability. We start by proving the second inequality \eqref{eq:ce_eps_2}: 
\begin{claim}\label{cl:eps_second}
\begin{align}
     \Pr\Big(\E[T^R_i(t)] - T^R_i(t) \ge \frac{1}{2} c k^{-\frac{2}{3}} (\ln(n))^{\frac{1}{3}} n^{\frac{-1}{3}} t \Big) \leq \frac{1}{n^4}
\end{align}
\end{claim}
\begin{proof}
We start by finding the $\E[T^R_i(t)]$: 
\begin{align*}
    \E[T^R_i(t)] & = \sum_{s=1}^t \frac{c k^{\frac{1}{3}} (\ln(n))^{\frac{1}{3}} n^{\frac{-1}{3}}}{k} \\ 
                 & = c k^{\frac{-2}{3}} (\ln(n))^{\frac{1}{3}} n^{\frac{-1}{3}} t 
\end{align*}
Now we apply a tail inequality of Theorem \ref{th:cherbound}:
\begin{align*}
    \Pr\Big(T^R_i(t) \leq \frac{1}{2}  \E[T^R_i(t)]\Big) & \leq \exp{(-\frac{1}{8} \E[T^R_i(t)])} \\ 
    & = \exp{(-\frac{1}{8} c k^{\frac{-2}{3}} (\ln(n))^{\frac{1}{3}} n^{\frac{-1}{3}} t)} \\ 
    & \leq \exp(-4 \ln(n)) && \text{(since $t \ge n^{\frac{2}{3}} k^{\frac{1}{3}} (\ln(n))^{\frac{1}{3}}$)}  \\
    & = \frac{1}{n^4}
\end{align*}
Now we can prove that the $\ce$ happens with high probability: 
\begin{claim}\label{cl:cleanevent_greedy}
\begin{align*}
    \Pr(\ce) \ge 1-\frac{3}{n^2}
\end{align*}
\end{claim}
\begin{proof}
Note that for the first condition in the clean event \eqref{eq:ce_eps_1} we can show that it fails for a given arm and round with probability at most $\frac{2}{n^4}$ using the same proof of Claim \ref{cl:ucb_clean_event} for UCB. Now consider some arm $i \in \arms$ and round $t  \ge n^{\frac{2}{3}} k^{\frac{1}{3}} (\ln(n))^{\frac{1}{3}}$, then by Claim \ref{cl:eps_second} we have the probability that inequality \eqref{eq:ce_eps_2} is upper bounded by $\frac{1}{n^4}$. Therefore, we can bound the $\be$ event (which is the complement of the $\ce$) using the union bound as follows:
\begin{align*}
    \Pr(\be) \leq \frac{3}{n^4} k n \leq \frac{3}{n^2}
\end{align*}
Therefore, we have $\Pr(\ce)\ge 1-\frac{3}{n^2}$. 
\end{proof}
% We use the same clean event used that was used in the regret bound of $\epsilon$-greedy (Theorem \ref{th:eps_greedy_regretbound}). We recall Claim \ref{cl:cleanevent_greedy}: 
% \begin{align*}
%     \Pr(\ce) \ge 1 - \frac{3}{n^2} 
% \end{align*}
Now we have the same claim as Claim \ref{cl:pulls_ce_lb} which we obtain through a similar proof: 
\begin{claim}\label{cl:pulls_ce_lb_eps_greedy}
\begin{align*}
    \E[T_i(n)|\ce] \ge \alpha n - \frac{3}{2} 
\end{align*}
\end{claim}
Now we note that for any given realization in $\epsilon$-greedy the total number of pulls $T_i(n)$ for arm $i$ can be written as $T_i(n)=T^E_i(n)+T^R_i(n)$ where $T^E_i(n)$ and $T^R_i(n)$ are the pulls from exploitation (when $s_t=0$) and exploration (when $s_t=1$), respectively. We use the following facts: 
\begin{fact}
\begin{align}
\E[T^R_i(n)|\ce] = 4 c k^{\frac{-2}{3}} (\ln(n))^{\frac{1}{3}} n^{\frac{2}{3}} = o(n)
\end{align}
\end{fact}
\begin{proof}
\begin{align*}
    \E[T^R_i(n)] & = \sum_{s=1}^n \frac{c k^{\frac{1}{3}} (\ln(n))^{\frac{1}{3}} n^{\frac{-1}{3}}}{k} \\ 
                 & = c k^{\frac{-2}{3}} (\ln(n))^{\frac{1}{3}} n^{\frac{2}{3}}  
\end{align*}         
Further, we have: 
\begin{align*}
    \E[T^R_i(n)|\ce]  & = \frac{\E[T^R_i(n)] -\E[T^R_i(n)|\be] \Pr(\be) }{\Pr(\ce)}  \\
                      & \leq \frac{\E[T^R_i(n)]-0}{\Pr(\ce)} \\ 
                      & \leq \frac{\E[T^R_i(n)]-0}{1-\frac{3}{n^2}} \\ 
                      & \leq 4 \E[T^R_i(n)] && \tag{$1-\frac{3}{n^2} \ge 1-\frac{3}{2^2}=\frac{1}{4}$ since $n \ge 2$} \\ 
                      & = 4 c k^{\frac{-2}{3}} (\ln(n))^{\frac{1}{3}} n^{\frac{2}{3}} \\ 
                      & = o(n) 
\end{align*}   
\end{proof}

This leads to the following lower bound on $\E[T^E_i(n)|\ce]$: 
\begin{claim}
\begin{align*}
    \E[T^E_i(n)|\ce] \ge \alpha n - o(n) 
\end{align*}
\end{claim}
\begin{proof}
First we start with the fact that $\E[T_i(n)] = \E[T^E_i(n)]+\E[T^R_i(n)]$ and therefore $\E[T^E_i(n)] = \E[T_i(n)]-\E[T^R_i(n)]$. Now have: 
\begin{align*}
    \E[T^E_i(n)]  = \Pr(\ce) \E[T^E_i(n)|\ce] + \Pr(\be) \E[T^E_i(n)|\be] 
\end{align*}
\begin{align*}
    \text{Therefore,} \ \ \E[T^E_i(n)|\ce] & \ge \frac{\E[T^E_i(n)] - \frac{3}{n^3} n }{\Pr(\ce)} \\ 
    & \ge \E[T^E_i(n)]  - \frac{3}{n} \\ 
    & = \alpha n -o(n)
\end{align*}
\end{proof}
Under the $\ce$ we have $\forall i \in \arms, \forall t \in [n]: |\meanit{t}-\mu_i| \leq \sqrt{\frac{2\ln(n)}{T_i(t)}} \text{ and } \forall i \in \arms, \forall t \ge n^{\frac{2}{3}} k^{\frac{1}{3}} (\ln(n))^{\frac{1}{3}}: T^R_i(t) \ge \frac{c}{2} t^{\frac{2}{3}} k^{-\frac{2}{3}} (\ln(n))^{\frac{1}{3}}$, it follows by the first condition of the $\ce$ that: 
\begin{align}
    \forall t \in [n]: \nonumber \\ 
    & \estmean{i}(t) \leq \mu_i + \sqrt{\frac{2\ln(n)}{T_i(t)}} + \frac{\sum_{s=1}^t c_i(s)}{T_i(t)} \label{eq:epsi} \\ 
    & \estmean{\honestarm}(t) \ge \honestmean - \sqrt{\frac{2\ln(n)}{T_{\honestarm}(t)}} \label{eq:epsh}
\end{align}
Consider some realization under the $\ce$ where round $t_0$ is the last \emph{exploitation} round where arm $i$ is pulled. It follows that: 
\begin{align*}
    \estmean{i}(t_0-1) & \ge \estmean{\honestarm}(t_0-1) \\
    \mu_i + \sqrt{\frac{2\ln(n)}{T_i(t_0-1)}} + \frac{\sum_{s=1}^{t_0-1} c_i(s)}{T_i(t_0-1)} & \ge \honestmean  - \sqrt{\frac{2\ln(n)}{T_{\honestarm}(t_0-1)}} \ \ \ \ \ \ \tag{By inequalities \eqref{eq:epsi} and \eqref{eq:epsh}}
\end{align*}
This is equivalent to
\begin{align}
    \sum_{s=1}^{t_0-1} c_i(s) & \ge (\honestgapi) T_i(t_0-1) - \sqrt{2\ln(n) T_i(t_0-1)} - T_i(t_0-1) \sqrt{\frac{2\ln(n)} {T_{\honestarm}(t_0-1)}} \nonumber \\ 
    & \ge (\honestgapi) T_i(t_0-1) - \sqrt{2\ln(n) T_i(t_0-1)} - o(n) \label{eq:bound_cost_sum}
\end{align}
where in the last step we have $T_i(t_0-1) \sqrt{\frac{2\ln(n)} {T_{\honestarm}(t_0-1)}}=o(n)$. This can be proved as follows. First, if $T_i(t_0-1) < n^{\frac{2}{3}} k^{\frac{1}{3}} (\ln(n))^{\frac{1}{3}}$ then we would have the following:
\begin{align*}
    T_i(t_0-1) \sqrt{\frac{2\ln(n)} {T_{\honestarm}(t_0-1)}} \leq \frac{ \sqrt{2} n^{\frac{2}{3}} k^{\frac{1}{3}} (\ln(n))^{\frac{5}{6}} }{1} = o(n) 
\end{align*}
If instead $T_i(t_0-1) \ge n^{\frac{2}{3}} k^{\frac{1}{3}} (\ln(n))^{\frac{1}{3}}$, then it must be that $t_0 \ge t_0 - 1 \ge T_i(t_0-1) \ge n^{\frac{2}{3}} k^{\frac{1}{3}} (\ln(n))^{\frac{1}{3}}$ and therefore since we are under the $\ce$ we must also have $T_{\honestarm}(t_0-1) \ge \frac{c}{4} k^{-\frac{2}{3}} (\ln(n))^{\frac{1}{3}} n^{\frac{-1}{3}} t_0$. Accordingly, we have:  
\begin{align*}
    T_i(t_0-1) \sqrt{\frac{2\ln(n)} {T_{\honestarm}(t_0-1)}} & \leq \frac{2\sqrt{2}}{\sqrt{c}} k^{\frac{1}{3}} (\ln(n))^{\frac{1}{3}} n^\frac{1}{6} \frac{T_i(t_0-1)}{\sqrt{t_0}} \\ 
    & \leq c' n^{\frac{5}{6}} (\ln(n))^\frac{1}{6} && \text{(where $c'$ is some constant)} \\ 
    & = o(n) 
\end{align*}
Note in the above that we invoked the fact that note that the number of arms $k$ is a constant. 

Before we go back to lower bounding the effort, note that $C_i(n)=\sum_{s=1}^{t_0-1} c_i(s) + c_i(t_0) + C^R_i(t_0 + 1;n)$ where $C^R_i(t_0 + 1;n)$ is the total effort spent in the remaining exploration rounds where $i$ is pulled. The following upper bound follows: 
\begin{align}
    C_i(n) & = \sum_{s=1}^{t_0-1} c_i(s) + c_i(t_0) + C^R_i(t_0 + 1;n) \nonumber \\ 
           & \ge \sum_{s=1}^{t_0-1} c_i(s) - 1 - T^R_i(n)  \label{eq:cost_bound_tot}
\end{align}
Furthermore, we note the following: 
\begin{align}
    T_i(t_0-1) & = T^E_i(t_0-1) + T^R_i(t_0-1) \nonumber \\  
               & = T^E_i(n) - 1 + T^R_i(t_0-1) \nonumber \\   
               & \ge T^E_i(n) - 1 \label{eq:bound_T_E}
\end{align}
Note further that: 
\begin{align}
    T_i(t_0-1) \leq T_i(n) \label{eq:bound_T_n}
\end{align}
If $(\honestgapi) \ge 0 $, then using Inequality \eqref{eq:bound_cost_sum}  in \eqref{eq:cost_bound_tot} leads to: 
\begin{align*}
    C_i(n) & \ge (\honestgapi) T_i(t_0-1) - \sqrt{2\ln(n) T_i(t_0-1)} - o(n) - 1 - T^R_i(n)  \\ 
           & \ge (\honestgapi) T^E_i(n)   -   \sqrt{2\ln(n) T_i(n)} - T^R_i(n) - o(n)  \ \ \ \text{(using Inequality \eqref{eq:bound_T_E})}
\end{align*}
Taking the expectation conditioned on the $\ce$ we have: 
\begin{align*}
    \E[C_i(n)|\ce] & \ge (\honestgapi) \E[T^E_i(n)|\ce] - \sqrt{2\ln(n) \E[T_i(n)|\ce]} \\
    & - \E[T^R_i(n)|\ce] - o(n)   \\
    & \ge \alpha n (\honestgapi) -o(n) 
\end{align*}
If $(\honestgapi) < 0 $, then using Inequality \eqref{eq:bound_T_n} we get: 
\begin{align*}
    C_i(n) & \ge (\honestgapi)  T_i(n)- \sqrt{2\ln(n)  T_i(n)} - o(n) - 1 - T^R_i(n)  \\ 
\end{align*}
Therefore, we have  
\begin{align*}
    \E[C_i(n)|\ce]  \ge & (\honestgapi) \E[T_i(n)|\ce] - \sqrt{2\ln(n) \E[T_i(n)|\ce]} \\ 
    & - \E[T^R_i(n)|\ce] - o(n)   \\
    \ge & \alpha n (\honestgapi) -o(n) 
\end{align*}
We get the same lower bound for $(\honestgapi)\ge 0$ and $(\honestgapi)<0$.We finally bound the total expected effort: 
\begin{align*}
    \E[C_i(n)] & = \E[C_i(n)|\ce] \Pr(\ce) + \E[C_i(n)|\be] \Pr(\be) \\ 
               & \ge \E[C_i(n)|\ce] \cdot \Big(1-\frac{3}{n^2}\Big) +  (-n) \cdot \frac{3}{n^2} \\ 
               & \ge (\honestgapi) (\alpha n-1) - o(n) \\ 
               & \ge (\honestgapi) \alpha n - o(n) 
\end{align*}
\end{proof}

We restate the next theorem and give its proof:
\ucbrobust*
\begin{proof}
The proofs follows from the proofs of Theorems \ref{th:ucbsa} for UCB and \ref{th:epsgreedysa} for $\epsilon$-greedy above.
\end{proof}

% \begin{oneshot}{theorem~\ref{Th:wet}}
% Iberian precipitation primarily hits the flatlands.
% \end{oneshot}

We restate the next theorem and prove it:
\domstra*
\begin{proof}
Consider two arms $\{a_1,a_2\}$ with means $\mu_1,\mu_2$ such that $1> \mu_1 > \mu_2$ and $M_1=M_2=1$. Suppose arm 2 follows the strategy $S_2$ of not changing its rewards at all. Arm 1 has two strategy options $S^*_1$ that always gives $1$ and $S_1$ that just takes the reward without changing it. Then we have the following:
\begin{align*}
    u_1(S^*_1,S_2) = \E[T_1(n)-C_1(n)] = \E[T_1(n)]-(1-\mu_1)\E[T_1(n)] = \mu_1 \E[T_1(n)] \leq \mu_1 n 
\end{align*}
Also, we have 
\begin{align*}
    u_1(S_1,S_2) = \E[T_1(n)] \ge n-o(n) 
\end{align*}
Therefore, $u_1(S_1,S_2) = u_1(S^*_1,S_2) + (1-\mu_1) n -o(n)$, therefore $S^*_1$ is not a dominant strategy.  
\end{proof}

We state the following fact and give its proof: 
% \begin{repfact}{factused}
% If a MAB algorithm is monotonic, then given strategy profile $S=(S_i,S_{-i})$ if $i$ deviates to $\Sbestheti{i}$ we have:
% \begin{flalign}
%      & \forall j \neq i , p\ge 0:  \Pr\big(T_j(n,\Sbestheti{i},S_{-i})\ge p \big) \leq \Pr\big(T_j(n,S_i,S_{-i})\ge p \big) \label{eq:mon_imp_1} \\ 
%      & \forall j \neq i:  \E[T_j(n,\Sbestheti{i},S_{-i})] \leq \E[T_j(n,S_i,S_{-i})] \label{eq:mon_imp_2}\\ 
%      & \E[\totpulls{i}(n,S^*_i,S_{-i})]\ge \E[\totpulls{i}(n,S_i,S_{-i})]  \label{eq:mon_imp_3}
% \end{flalign}
% \end{repfact}

\begin{restatable}{fact}{factused}\label{fact:mono_implications}
If a MAB algorithm is monotonic, then given strategy profile $S=(S_i,S_{-i})$ if $i$ deviates to $\Sbestheti{i}$ we have:
\begin{flalign}
     & \forall j \neq i , p\ge 0: \nonumber \\ 
     & \Pr\big(T_j(n,\Sbestheti{i},S_{-i})\ge p \big) \leq \Pr\big(T_j(n,S_i,S_{-i})\ge p \big) \label{eq:mon_imp_1} \\ 
     & \forall j \neq i:  \E[T_j(n,\Sbestheti{i},S_{-i})] \leq \E[T_j(n,S_i,S_{-i})] \label{eq:mon_imp_2}\\ 
     & \E[\totpulls{i}(n,S^*_i,S_{-i})]\ge \E[\totpulls{i}(n,S_i,S_{-i})]  \label{eq:mon_imp_3}
\end{flalign}
\end{restatable}
\begin{proof}
The first and second property follow immediately since for every reward and strategy seed we have $\totpulls{j}(n,S^*_i,S_{-i}) \leq \totpulls{j}(n,S_i,S_{-i})$. We prove the third property: 
\begin{align*}
    & n = \E[\totpulls{i}(n,S_i,S_{-i})] + \sum_{j \neq i} \E[\totpulls{j}(n,S_i,S_{-i})] = \E[\totpulls{i}(n,S^*_i,S_{-i})] + \sum_{j \neq i} \E[\totpulls{j}(n,S^*_i,S_{-i})]. 
\end{align*}
Rearranging the equation, we obtain
\begin{align*}
    \E[\totpulls{i}(n,S^*_i,S_{-i}] - \E[\totpulls{i}(n,S_i,S_{-i})] =  \sum_{j \neq i} \Big( \E[\totpulls{j}(n,S_i,S_{-i})] - \E[\totpulls{j}(n,S^*_i,S_{-i})] \Big) \ge 0,
\end{align*}
where for the inequality, we invoke the second property $\E[T_j(n,S_i,S_{-i})] \ge \E[T_j(n,\Sbestheti{i},S_{-i})]$ for $j \neq i$. 
\end{proof}

Now we prove that UCB is monotonic 
\begin{theorem}\label{th:ucbmono}
UCB is monotonic. 
\end{theorem}
\begin{proof}
Consider the strategy profiles $S=(S_i,S_{-i})$ and $S'=(S^*_i,S_{-i})$. Note that since we are in the blind information model, every arm only sees its own history and hence its strategy is dependent only on its own history. Further, under strategy $S^*_i$ the reward at the $\ell^{th}$ pull given a specific history $h^{\ell-1}_i$ for arm $i$ is $M_i$. 

Consider both strategy profiles under $S$ and $S'$ under the same reward realizations and seeds: (1) Any arm $j \in [k]$ realizes the same reward when pulled the same number of times. (2) Every arm $j \in [k], j\neq i$ has the same seed for its (possibly) randomized strategy. We will show that under these conditions, it must be that $\totpulls{j}(n,S') \leq \totpulls{j}(n,S')$. We denote the sequence of pulls given a profile $S$ up to and including $t$ by $\pullseq(t,S)(-i)$ and we denote the same sequence with the pulls of arm $i$ excluded. We start with the following claim which closely follows Lemma B.2 from \cite{feng2020intrinsic}: 
\begin{claim}
Given the same reward realizations and seeds if $t\ge k$, then $\pullseq(t,S')(-i)$ is a sub-sequence of $\pullseq(t,S)(-i)$\footnote{If sequence $A$ is a sub-sequence of sequence $B$, then either $A=B$ or $A$ is a proper sub-sequence of $B$ (having less elements).}. 
\end{claim}
\begin{proof}
The proof is by induction. For the base case we set $t=k$, then the sub-sequences are identical $\pullseq(t,S')(-i)=\pullseq(t:t',S)(-i)$. 

Now at a given $t+1$, by the inductive hypothesis $\pullseq(t,S')(-i)$ is a sub-sequence of $\pullseq(t,S)(-i)$. This means that either $\pullseq(t,S')(-i)=\pullseq(t,S)(-i)$ or that $\pullseq(t,S')(-i)$ is a proper sub-sequence of $\pullseq(t,S)(-i)$. If $\pullseq(t,S')(-i)=\pullseq(t,S)(-i)$, then $\forall j\in [k], j\neq i$ we have $\ucbi{j}(t,S')=\ucbi{j}(t,S)$ and $\ucbi{i}(t,S')\ge \ucbi{i}(t,S)$. Accordingly, it is impossible for an arm $j\neq i$ to be pulled for $S'$ and for arm $i$ to pulled for $S$ at round $t+1$. The possibilities are: (1) both pull the same arm $j \neq i$ since $\ucbi{j}(t,S')=\ucbi{j}(t,S)$. (2) both pull $i$. (3) arm $i$ is pulled for $S'$ but an arm $j \neq i$ is pulled for $S$. Possibilities (1) and (2) lead to $\pullseq(t+1,S')(-i)=\pullseq(t+1,S)(-i)$ whereas (3) leads to $\pullseq(t+1,S')(-i)$ being a proper sub-sequence of $\pullseq(t+1,S)(-i)$.

If $\pullseq(t,S')(-i)$ is a proper sub-sequence of $\pullseq(t,S)(-i)$, this implies that $|\pullseq(t,S')(-i)|=q < |\pullseq(t,S)(-i)|$. Therefore, for all $j \neq i$, $\ucbi{j}(t,S')=\ucbi{j}(t(q),S)$ where $\ucbi{j}(t(q),S)$ is the UCB index for arm $j$ in the round corresponding to the $q^{th}$ element in the sequence. It follows that if at round $t+1$ arm $i$ is not pulled for $S'$, then $S'$ will pull the arm at element $q+1$ of $\pullseq(t,S)(-i)$ and therefore $\pullseq(t+1,S')(-i)$ would still be a sub-sequence of $\pullseq(t+1,S)(-i)$ regardless of the arm pulled for $S$. If instead arm $i$ is pulled for $S'$ then $\pullseq(t+1,S')(-i)$ is a proper sub-sequence of $\pullseq(t+1,S)(-i)$ regardless of the arm pulled for $S$. 
\end{proof}
It follows immediately by the above claim that $\totpulls{j}(n,S') \leq \totpulls{j}(n,S)$. Therefore, UCB is monotonic. 
\end{proof}

Now we prove that UCB is FATA:
\begin{theorem}\label{th:ucbfata}
UCB is FATA. 
\end{theorem}
\begin{proof}
Suppose we have a collection of arms $\armsr$ where each arm $i \in \armsr$ gives $r^*=\max_{t\in [n]} r_t$  whenever pulled. Then $\forall i,i' \in \armsr$, we have at $t=k$ that $\totpulls{i}(k)=\totpulls{i'}(k)$. 

Now, for any round $t>k$ suppose by contradiction that $\totpulls{i}(t)\ge\totpulls{i'}(t)+2$, then there must be a round $t_1<t$ such that $\totpulls{i}(t_1) = \totpulls{i'}(t)+1$. There must also be a round $t_1 < t_2 \leq t$ where arm $i$ is pulled again to make $\totpulls{i}(t)=\totpulls{i'}(t)+2$ but at $t_2-1$, we have: 
\begin{align*}
    \ucbi{i}(t_2-1) = r^* + \sqrt{\frac{2\ln(n)}{\totpulls{i}(t_1)}} < r^* + \sqrt{\frac{2\ln(n)}{\totpulls{i'}(t)}} \leq r^* + \sqrt{\frac{2\ln(n)}{\totpulls{i'}(t_2-1)}} = \ucbi{i'}(t_2-1) 
 \end{align*}
since $\ucbi{i'}(t_2-1)  > \ucbi{i}(t_2-1)$, it cannot be the case that arm $i$ is pulled at round $t_2$ which leads to a contradiction. Therefore, it must be the case that $\totpulls{i}(t)<\totpulls{i'}(t)+2$. By exchanging the indices, we also reach the conclusion that $\totpulls{i}(t)>\totpulls{i'}(t)-2$. Therefore, $|\totpulls{i}(t)-\totpulls{i'}(t)| \leq 1$. This implies that $\forall i,i' \in \armsr, \ |\E[\totpulls{i}(n)] -\E[\totpulls{i'}(n)]|\leq 1 =o(n)$. 
\end{proof}

Now we prove that $\epsilon$-greedy is monotonic:
\begin{theorem}\label{th:ueps_monotonic}
$\epsilon$-Greedy is monotonic. 
\end{theorem}
\begin{proof}
Consider the strategy profiles $S=(S_i,S_{-i})$ and $S'=(S^*_i,S_{-i})$. Note that since we are in the blind information model, every arm only sees its own history and hence its strategy is dependent only on its own history. Further, under strategy $S^*_i$ the reward at the $\ell^{th}$ pull given a specific history $h^{\ell-1}_i$ for arm $i$ is always $M_i$. 

Consider both strategy profiles under $S$ and $S'$ under the same reward seed (realizations) and strategy seeds: (1) Any arm $j \in [k]$ realizes the same reward when pulled the same number of times. (2) Every arm $j \in [k], j\neq i$ has the same seed for its (possibly) randomized strategy. (3) The $\epsilon-$Greedy algorithm has the same seed, this implies that the algorithm would have the same rounds for exploration and exploitation. Further, in  an exploration round the same arms would be chosen and in an exploitation round if the same set of arms have the maximum mean then the same arm would be chosen.

We will show that it must be that $\totpulls{j}(n,S') \leq \totpulls{j}(n,S)$. We denote the sequence of pulls given a profile $S$ up to and including $t$ with arm $i$ excluded by $\pullseq(t,S)(-i)$. We start with the following claim: 
\begin{claim}
Given the same reward realizations and seeds if $t\ge k$, then $\pullseq(t,S')(-i)$ is a sub-sequence of $\pullseq(t,S)(-i)$\footnote{If sequence $A$ is a sub-sequence of sequence $B$, then either $A=B$ or $A$ is a proper sub-sequence of $B$ (having less elements).}. 
\end{claim}
\begin{proof}
The proof is by induction. For the base case we set $t=k$, then the sub-sequences are identical $\pullseq(t,S')(-i)=\pullseq(t,S)(-i)$. 

Now at a given $t+1$, by the inductive hypothesis $\pullseq(t,S')(-i)$ is a sub-sequence of $\pullseq(t,S)(-i)$. This means that either $\pullseq(t,S')(-i)=\pullseq(t,S)(-i)$ or that $\pullseq(t,S')(-i)$ is a proper sub-sequence of $\pullseq(t,S)(-i)$. If $\pullseq(t,S')(-i)=\pullseq(t,S)(-i)$, then round $t+1$ is either an exploration or exploitation round. If it is an exploration round, this since we have the same seed for $\epsilon-$Greedy, then the same arm will be pulled under $S'$ and $S$. Therefore, we would have $\pullseq(t+1,S')(-i)=\pullseq(t+1,S)(-i)$. If $t+1$ is an exploitation round, then since $\pullseq(t,S')(-i)=\pullseq(t,S)(-i)$ and the arms have the same rewards and seeds, then $\emp{j}(t,S')=\emp{j}(t,S)$, but $\emp{i}(t,S') \ge \emp{i}(t,S)$. Therefore, if $\emp{i}(t,S')<\max_{j \neq i} \emp{j}(t,S')$, then since we have the same $\epsilon-$Greedy, the same arm would be pulled and $\pullseq(t+1,S')(-i)=\pullseq(t+1,S)(-i)$. However, if $\emp{i}(t,S')\ge\max_{j \neq i} \emp{j}(t,S')$, then dependent on the seed of $\epsilon-$Greedy either arm $i$ would be pulled under $S'$ or the same arm $j\neq i$ would be pulled in case there is an arm $j \neq i$ which also has the maximum empirical mean. Therefore, either $\pullseq(t+1,S')(-i)=\pullseq(t+1,S)(-i)$ or $\pullseq(t+1,S')(-i)$ is a proper sub-sequence of $\pullseq(t+1,S)(-i)$. 

If $\pullseq(t,S')(-i)$ is a proper sub-sequence of $\pullseq(t,S)(-i)$, this implies that $|\pullseq(t,S')(-i)|=q < |\pullseq(t,S)(-i)|$. Therefore, for all $j \neq i$, $\empi{j}(t,S')=\empi{j}(t(q),S)$ where $\empi{j}(t(q),S)$ is the empirical mean for arm $j$ in the round corresponding to the $q^{th}$ element in the sequence. If round $t+1$ is an exploitation round, then if arm $i$ is not pulled for $S'$, then $S'$ will pull the arm at element $q+1$ of $\pullseq(t,S)(-i)$ and therefore $\pullseq(t+1,S')(-i)$ would still be a sub-sequence of $\pullseq(t+1,S)(-i)$ regardless of the arm pulled for $S$. If instead arm $i$ is pulled for $S'$ then $\pullseq(t+1,S')(-i)$ is a proper sub-sequence of $\pullseq(t+1,S)(-i)$ regardless of the arm pulled for $S$. In the case where round $t+1$ is an exploration round, then the same arm would be pulled and therefore $\pullseq(t+1,S')(-i)$ would be a proper sub-sequence of $\pullseq(t+1,S)(-i)$.
\end{proof}
It follows by the above claim that $\totpulls{j}(n,S')\leq \totpulls{j}(n,S)$. Therefore, $\epsilon-$Greedy is monotonic. 
\end{proof}

We now prove the FATA property for $\epsilon$-greedy:  
\begin{theorem}\label{th:epsfata}
$\epsilon$-Greedy is FATA. 
\end{theorem}
\begin{proof}
We note that for at $t=k$ each arm has been pulled exactly once and that $\forall t\ge k$, we have $\forall i,i' \in \armsr: \hat{\mu}_i(t) = \hat{\mu}_{i'}(t)$ by definition of $\armsr$. Therefore, since when $s_t=0$ (exploitation), then algorithm picks among the top arms uniformly at random we have: 
\begin{align*}
    \forall t> k: & \Pr[I_t=i|s_t=0] = \Pr[I_t=i'|s_t=0] \\
    \implies & \E[T^E_i(n)] = \sum_{t=1}^n \Pr[I_t=i|s_t=0] \Pr[s_t=0] = \sum_{t=1}^n \Pr[I_t=i'|s_t=0] \Pr[s_t=0] = \E[T^E_{i'}(n)]
\end{align*}
Since $\forall i,j \in \arms: \E[T^R_i(n)]=\E[T^R_j(n)]$, then $\E[T^R_i(n)]=\E[T^R_{i'}(n)]$ and therefore $\E[T_i(n)]=\E[T_{i'}(n)]$ and the FATA property is satisfied. 
\end{proof}

We restate the next theorem and give it proof: 
\ucbrpi*
\begin{proof} 
We proved that UCB and $\epsilon$-greedy are sharply adaptive in Theorem \ref{th:UCB1_robust}. In the appendix, Theorems (\ref{th:ucbmono},\ref{th:ucbfata}) for UCB and Theorems (\ref{th:ueps_monotonic},\ref{th:epsfata}) for $\epsilon$-greedy prove that they satisfy monotonicity and FATA. Hence UCB and $\epsilon$-greedy are SAMF algorithms. 
\end{proof}

We restate the next theorem and give it proof: 
\nobadequlib*
\begin{proof}
The idea of the proof to show that if the revenue $P(n,S)\leq \alpha \maxall n +o(n)$ then there must be a linear proportion of the horizon where the performance is sub-optimal. There must also exist an arm $i \in \armstop$ which receives at most $\frac{1}{\bk}$ of that horizon and therefore if $i$ deviates to top performance it can get higher utility if Condition \ref{cond:cond_good_revenue} is satisfied.

We start with the following claim which shows that top performance is not played by any arm $i \in \armstop$.
\begin{claim}\label{cl:first_cl}
If $P(n,S)\leq \alpha \maxall n + o(n)$ where $\alpha <1$ for some strategy profile $S=(S_1,\dots,S_k)$, then it cannot be the case that there exists an arm $i \in \armstop$ such that $S_i=S^*_i$.
\end{claim}
\begin{proof}

By contradiction, if that was the case, then the MAB instance can be reduced to an instance where the honest agent with the maximum mean $\honestarm$ has $\honestmean=\maxall$. Since a sharply adaptive algorithm always has $P(n,S) \ge \maxall n -o(n)$ then by Theorem \ref{th:sr_always_sb_regret} we have a contradiction. 
\end{proof}

First, for a given arm $j$ define the indicator random variable $\pc{j}(p,S)$ where $\pc{j}(p,S)=1$ if arm $j$ has under strategy profile $S$ been pulled for the $p^{\text{th}}$ time and is $0$ otherwise. It follows by definition that we have:
\begin{fact}\label{fact:pc_facts}
\begin{align}
                     & \Pr\Big(\pc{j}(p,S)=1\Big) = \Pr\Big(T_j(n,S)\ge p\Big)  \\
                     & \sum_{p=1}^n \E[\pc{j}(p,S)] = \E[T_j(n,S)] \label{eq:pc_to_tn} \\ 
    \forall j\neq i: &  \Pr\Big(\pc{j}(p,\Sbestheti{i},S_{-i})=1 \Big) \leq \Pr\Big(\pc{j}(p,S_i,S_{-i})=1 \Big) \label{eq:pc_less}
\end{align}
\end{fact}
\begin{proof}
The first property follows by definition. 

The second property is a well-known Lemma (see, e.g. Lemma 2.9 in \cite{mitzenmacher2017probability}). We add the proof for completeness. 
\begin{align*}
    \sum_{p=1}^n \E[\pc{j}(p,S)] & = \sum_{p=1}^n \Pr\Big(T_j(n,S)\ge p\Big) \\ 
                               & = \sum_{p=1}^n \sum_{t=p}^n \Pr\Big(T_j(n,S) = t \Big) \\ 
                               & = \sum_{t=1}^n \sum_{p=1}^t \Pr\Big(T_j(n,S) = t \Big) \\ 
                               & = \sum_{t=1}^n t \Pr\Big(T_j(n,S) = t \Big) \\ 
                               & = \E[T_i(n,S)]
\end{align*}

The third property follows since monotonicity implies that if arm $i$ deviates from $S_i$ to $\Sbestheti{i}$ then for any arm $j \neq i$ we have $\Pr\Big(T_j(n,\Sbestheti{i},S_{-i})\ge p \Big) \leq \Pr\Big(T_j(n,S_i,S_{-i})\ge p \Big)$ (see inequality (\ref{eq:mon_imp_1}) in Fact \ref{fact:mono_implications}). 
\end{proof}

Now, define $\tildr_j(p,S)$ to be the random variable for the modified (final) reward (i.e., sampled reward + effort) which arm $j$ gives when it is pulled for $p^{\text{th}}$ time under strategy profile $S$. 
%Note that there is an abuse of notation, $\tildr_j(p)$ with the symbol $p$ refers to the reward arm $j$ gives for the $p^{\text{th}}$ pull whereas the previous notation $\tildr_j(t)$ with $t$ instead referred to the reward from $j$ for round $t$. 
Since we are in a blind observation model (see Section \ref{sec:model}) where the arms are only aware of their own history and not the round value or the other arms' histories, it follows that if a subset of arms other than $j$ were to deviate from $S_{-j}$ to $S'_{-j}$ then we would have: 
\begin{align}\label{eq:reward_inv}
    \E[\tildr_j(p,S_j,S_{-j})|\pc{j}(p,S_j,S_{-j})=1] & = \E[\tildr_j(p,S_j,S'_{-j})|\pc{j}(p,S_j,S'_{-j})=1]  
    \\ 
    & = \E[\tildr_j(p,S_j)|\pc{j}(p,S_j)=1] 
\end{align}
That is, the expected reward an arm $j$ gives for the $p^{\text{th}}$ pull (conditioned on the fact that the pull is successful) is dependent only on its own strategy $S_j$ . 

The ``final'' reward given by arm $j$ for the $p^{\text{th}}$ pull is by definition the following: 
\begin{align}\label{eq:reward_inv2}
    \E[\tildr_j(p,S_j, S_{-j})]  = \E[\tildr_j(p,S_j, S_{-j})|\pc{j}(p,S_j, S_{-j})=1] \Pr\Big(\pc{j}(p,S_j, S_{-j})=1 \Big)
\end{align}

For a given arm $j$ with strategy $S_j$, each pull value $p \in [n]$ has associated with it an expected reward value $\E[\tildr_j(p,S_j)|\pc{j}(p,S_j)=1]$. Therefore we can define for a given arm $j \in \arms$ two \emph{disjoint} subset of rounds $\poptj{j}(S_j)$ and $\psuboptj{j}(S_j)$ according to the value of $\E[\tildr_j(p,S_j)|\pc{j}(p,S_j)=1]$. Specifically, we have:
\begin{align}
    & \text{If $\E[\tildr_j(p,S_j)|\pc{j}(p,S_j)=1] \ge \maxall - o(1)$, then $p \in \poptj{j}(S_j)$}  \\ 
    & \text{Otherwise $p \in \psuboptj{j}(S_j)$} 
\end{align}
Note that by \eqref{eq:reward_inv} we have $\poptj{j}(S_j)=\poptj{j}(S_j,S_{-j})=\poptj{j}(S)$, similarly $\psuboptj{j}(S_j)=\psuboptj{j}(S_j,S_{-j})=\psuboptj{j}(S)$.
Note that according to the above if $p \in \psuboptj{j}(S_j)$, then $\E[\tildr_j(p,S)|\pc{j}(p,S)=1] \leq \beta \maxall +o(1)$ where $\beta<1$ is a constant.

We further define the following: 
\begin{align}
& \toptj{j}(S)  = \E[\sum_{p\in \poptj{j}} \pc{j}(p,S)] \\
& \tsuboptj{j}(S) = \E[\sum_{p\in \psuboptj{j}} \pc{j}(p,S)] 
\end{align}
%\SH{no need for conditioning?}
The above construction implies the following: 
\begin{align}
& |\poptj{j}(S)| + |\psuboptj{j}(S)| = n \\ 
& \E[\sum_{p\in \poptj{j}} \tildr_j(p,S)] \ge \maxall \toptj{j}(S)  - o(n) \label{eq:topt_def}\\
& \E[\sum_{p\in \psuboptj{j}} \tildr_j(p,S)] \leq \beta \maxall \tsuboptj{j}(S) + o(n) \ \ \text{where the constant $\beta <1$}
\end{align}
%\SH{no need for conditioning?}
Further, define $\topt(S) = \sum_{j \in \arms} \toptj{j}(S)$ and $\tsubopt(S)= \sum_{j \in \arms} \tsuboptj{j}(S)$.

For the given strategy profile $S$ with $P(n)=P(n,S)\leq \alpha \maxall n +o(n)$, we lower bound $\tsubopt(S)$: 
\begin{claim}\label{cl:tsubopt}
$\tsubopt(S) \ge (1-\alpha)n - o(n)$.
\end{claim}
\begin{proof} 
\begin{align*}
    P(n,S) & = \sum_{j \in \arms} \sum_{p=1}^n \E[\tildr_j(p,S)] \\ 
         & = \Big(\sum_{j \in \arms} \sum_{p \in \poptj{j}} \E[\tildr_j(p,S)] \Big) + \Big(\sum_{j \in \arms} \sum_{p \in \psuboptj{j}} \E[\tildr_j(p,S)] \Big) \\ 
\end{align*}
Therefore, we have
\begin{align*}
    \maxall \sum_{j \in \arms} \toptj{j}(S) - o(n) & \leq \Big(\sum_{j \in \arms} \sum_{p \in \poptj{j}} \E[\tildr_j(p,S)] \Big)  \ \ \ \ \ \tag{from \eqref{eq:topt_def}} \\
    & = P(n,S) - \Big(\sum_{j \in \arms} \sum_{p \in \psuboptj{j}} \E[\tildr_j(p,S)] \Big) \\ 
    & \leq P(n,S) - 0  \\
    & \leq \alpha \maxall n - o(n) \\ \\ 
    \text{Therefore,  }  \topt(S) = \sum_{j \in \arms} \toptj{j}(S)  &  \leq \alpha n + o(n)
\end{align*}
Note that by construction of $\poptj{j}(S)$ and $\psuboptj{j}(S)$ and \eqref{eq:pc_to_tn} in Fact \ref{fact:pc_facts}:
\begin{align*}
 \E[T_j(n,S)] = \toptj{j}(S) + \tsuboptj{j}(S)
\end{align*}
Further we always have: 
\begin{align*}
\sum_{j \in \arms}  \E[T_j(n,S)] = n 
\end{align*}
Therefore, it follows that:
\begin{align*}
    \tsubopt(S)   =  n - \sum_{j \in \arms} \toptj{j}(S) \ge (1-\alpha)n - o(n)
\end{align*} 
\end{proof}

Since there are $\bk$ many arms in $\armstop$, there must exist an arm $i \in \armstop$ such that $\tsuboptj{i}(S) \leq \frac{\tsubopt(S)}{\bk}$. We will consider $i$'s  deviation from $S_i$ to $\Sbestheti{i}$. Accordingly, the original strategy profile is $S=(S_i,S_{-i})$ and the deviation strategy profile is $S'=(\Sbestheti{i},S_{-i})$. Our specification of $i$ implies the following utility upper bound under $S$ for $i$:
\begin{align}
    u_i(S) & \leq \toptj{i}(S) (1+\mu_i-\maxall)  + \frac{\tsubopt(S)}{\bk} (1+\mu_i) + o(n)
\end{align}

We have the following claim: 
\begin{claim}\label{cl:jpulls_subreduc}
$\forall j \neq i: \toptj{j}(S') \leq \toptj{j}(S)$
\end{claim}
\begin{proof}
    For any $p \in \poptj{j}$, observe that
    \begin{align*}
        \Pr\Big(\pc{j}(p,\Sbestheti{i},S_{-i})=1 \Big) & \leq \Pr\Big(\pc{j}(p,S_i,S_{-i})=1 \Big),
    \end{align*}
    due to \eqref{eq:pc_less} in Fact \ref{fact:pc_facts}.
    Therefore, we have
\begin{align*}
    \toptj{j}(S') & = \sum_{p \in \poptj{j}} \Pr\Big(\pc{j}(p,\Sbestheti{i},S_{-i})=1 \Big) \\ 
    & \leq \sum_{p \in \poptj{j}} \Pr\Big(\pc{j}(p,S_i,S_{-i})=1 \Big) \\ 
    & = \toptj{j}(S) 
\end{align*}
\end{proof}

Now we introduce the following claim which essentially states that $ \tsuboptj{j}(S')$ vanishes $\forall j\neq i$. 
\begin{claim}\label{cl:jpulls_optreduc}
$\forall j \neq i: \tsuboptj{j}(S') =o(n)$
\end{claim}
\begin{proof}
Under the deviating strategy profile $S'$ arm $i$ follows $\Sbestheti{i}$, then by sharp adaptivity we must have:
\begin{align*}
     P(n,\Sbestheti{i},S_{-i}) = \maxall n - o(n)  
\end{align*}   
Therefore, by decomposing the rewards over the arms we have: 
\begin{align*}
     P(n,\Sbestheti{i},S_{-i}) & = \Big( \sum_{j' \neq j} \E[\sum_{p=1}^{n} \tildr_{j'}(p,\Sbestheti{i},S_{-i}) ] \Big) + \Big(\E[\sum_{p=1}^{n} \tildr_j(p,\Sbestheti{i},S_{-i}) ] \Big)  \\ 
    & \leq \maxall \Big((n-\big(\toptj{j}(S')+\tsuboptj{j}(S')\big)\Big) + \Big(\E[\sum_{p=1}^{n} \tildr_j(p,\Sbestheti{i},S_{-i})] \Big)  \\ 
    & \leq \maxall \Big((n-\big(\toptj{j}(S')+\tsuboptj{j}(S')\big)\Big) + \maxall \toptj{j}(S') + \beta' \maxall \tsuboptj{j}(S') + o(n) \\ 
    & = \maxall n + (\beta'-1) \maxall \tsuboptj{j}(S') + o(n)
\end{align*} 
Where in the above $\beta'$ is some constant and $\beta'<1$. Since  $P(n,\Sbestheti{i},S_{-i}) = \maxall n - o(n)$, then we have:
\begin{align}
\maxall n - o(n) & \leq \maxall n + (\beta'-1) \maxall \tsuboptj{j}(S') + o(n) \nonumber \\ 
\iff - o(n) & \leq (\beta'-1) \maxall \tsuboptj{j}(S') \\ 
\iff  \tsuboptj{j}(S') & \leq \frac{-1}{(\beta'-1)\maxall}  o(n) = o(n)   
\end{align} 
Note in the above that $\beta'<1$ and therefore $(\beta'-1)<0$.
\end{proof}

Now we have the following  claim which lower bounds the pulls of arm $i$ under the deviating strategy: 
\begin{claim}\label{cl:lower_bound_i_pulls}
$\E[T_i(n,S')] \ge \toptj{i}(S) + \tsubopt(S) - o(n)$.
\end{claim}
\begin{proof}
We start by upper bounding the pulls of the other arms: 
\begin{align*}
    \sum_{j \neq i} \E[T_j(n,S')] & \leq \sum_{j \neq i} \toptj{j}(S) + o(n) \ \ \ \tag{by Claims \ref{cl:jpulls_subreduc} and \ref{cl:jpulls_optreduc}} \\ 
    & = \topt(S) - \toptj{i}(S) + o(n) 
\end{align*}
Therefore, we have: 
\begin{align*}
     \E[T_i(n,S')] 
     & = 
     n - \sum_{j \neq i} \E[T_j(n,S')] 
     \\ 
     & \ge 
     n - \topt(S) + \toptj{i}(S) - o(n)
     \\ 
     & = 
     \toptj{i}(S) + \tsubopt(S) - o(n) 
\end{align*}    
\end{proof}

Now we can lower bound $i$'s utility under $S'$:
\begin{claim}
$u_i(S')\ge \toptj{i}(S) (1+\mu_i-\maxall)  + \tsubopt(S) (1+\mu_i-\maxall)   - o(n)$
\end{claim}
\begin{proof}
    Note that $u_i(S')$ can be expanded as follows.
    \begin{align*}
        u_i(S') & = \E[T_i(n,S')] - \E[C_i(n,S')] \\ 
            & = (1+\meani-\maxall) \E[T_i(n,S')] \ \ \ \tag{since arm $i$ follows $\Sbestheti{i}$}\\ 
            & \ge \toptj{i}(S) (1+\mu_i-\maxall)  + \tsubopt(S) (1+\mu_i-\maxall)   - o(n) \ \ \tag{by Claim \ref{cl:lower_bound_i_pulls}}
    \end{align*}
\end{proof}
Note that $\tsubopt(S)=\Omega(n)$ by Claim \ref{cl:tsubopt}. Now since Condition \ref{cond:cond_good_revenue} states that $\bk > \frac{2}{\min_{i \in \armstop} (1+\meani-\maxall)} (1+\epsilon)$, then we have $(1+\meani-\maxall) > \frac{2(1+\epsilon)}{\bk} = \frac{2\epsilon}{\bk} + \frac{2}{\bk} \ge \frac{2\epsilon}{\bk} + \frac{1+\mu_i}{\bk}$, since the original utility is $u_i(S) \leq \toptj{i}(S) (1+\mu_i-\maxall) + \frac{\tsubopt(S)}{\bk} (1+\mu_i)$. It follows that $\lim\limits_{n \xrightarrow[]{}\infty} \frac{u_i(S')}{u_i(S)}=\infty$. Therefore, from the above it easily follows that we do not have a partial asymptotic equilibrium.
\end{proof}

We restate the next theorem and give its proof:
\compinfequilib*
\begin{proof}
We begin with the following claim which upper bounds the number of pulls for all arms $i' \in \arms-\armsup$: 
\begin{claim}
Given strategy profile $\Sbesthet$, then for any strategy profile $\Sbestheto$ followed by arms $i' \in \arms-\armsup$, we have $\forall i' \in \arms-\armsup: \E[\totpulls{i'}(n,\Sbesthet,\Sbestheto)]=o(n)$. 
\end{claim}
\begin{proof}
Since every arm $ i \in \armsup$ follows strategy $\Sbestheti{i}$ of always giving $\maxall$ it follows that this instance is reducible to one where the maximum honest agent has a mean of $\honestmean=\maxall$. But by sharp adaptivity if arm $i'$ has $\E[\totpulls{i'}(n,\Sbesthet,\Sbestheto)] = \alpha n$, then it follows that $\E[\toteffort{i'}(n,\Sbesthet,\Sbestheto)] \ge (\maxall-\mu_{i'}) \alpha n -o(n)$. But since $i'$ can give a maximum value of $M_{i'}$ it follows that $\E[\toteffort{i'}(n,\Sbesthet,\Sbestheto)] \leq (M_{i'}-\mu_{i'}) \alpha n + o(n)= (\maxall-\mu_{i'}) \alpha n -(\maxall-M_{i'}) \alpha n + o(n)= (\maxall-\mu_{i'}) \alpha n  - \Omega(n)$. Therefore, regardless of the strategies followed by arms $i' \in \arms-\armsup$, they will always have $\E[\totpulls{i'}(n,\Sbesthet,\Sbestheto)]=o(n)$. 
% Fix any arm $i' \in \arms-\armsup$ and suppose it follows strategy $\Sbestahet{i'}$ of always giving its maximum possible reward value of $M_{i'}$. Therefore, the MAB instance has been reduced to one where $\forall i \in \armsup$ rewards are sampled from a deterministic distribution which always gives $\maxall$ and for $i' \in \arms-\armsup$ the reward is sampled from another deterministic distribution which always gives $M_i$. Therefore, the reward gap for $i'$ becomes $\Delta_{i'}=\maxall-M_{i'}$. Since arm $i'$ cannot generate a higher reward than $M_{i'}$, it follows that in this reduction that $C_{i'}=0$ and therefore by Fact (\ref{fact:sr_sub_pulls}) regardless of the strategy profile $S''$ followed by the other arms $j' \in \arms-\armsup, j'\neq i'$, we have $\E[\totpulls{i'}(n)|\Sbestahet{i'},\Sbesthet,S'']=o(n)$. 

% Since the algorithm is monotonic it follows that $i'$ can maximize its number of pulls by always the maximum possible reward, i.e. strategy $\Sbestahet{i'}$. It follows that $\forall i' \in \arms-\armsup$ we have $\E[\totpulls{i'}(n)|\Sbesthet,\Sbestheto]=o(n) \ , \forall \Sbestheto$.   
\end{proof}
\begin{lemma}
\begin{align}
   \forall i \in \armstop: \ \  u_i(S^*_i,S^*_{-i}) \ge (1+\meani-\maxall) \frac{n}{\bk} - o(n) 
\end{align}
\end{lemma}
\begin{proof}
By the above claim, it follows that only arms in $i \in \armsup$ can have $\E[\totpulls{i}(n)]=\Omega(n)$. Since all arms in $\armsup$ always give a reward of $\maxall$, then it follows by the FATA property that $\E[\totpulls{i}(n)] \ge \frac{n}{\bk}-o(n)$, therefore we have:
\begin{align}
        u_i(S^*_i,S^*_{-i}) = \E[\totpulls{i}(n)] - \E[\toteffort{i}(n)] = \E[\totpulls{i}(n)] (1-(M_i-\meani)) \ge (1+\meani-\maxall) \frac{n}{\bk} - o(n) 
\end{align}
\end{proof}
We now upper bound the utility of an arm $i$ given $S^*_{-i}$: 
\begin{lemma}\label{lemma:pi_util_upper_bound}
\begin{align}
    \forall i \in \armstop: \ \    u_i(S_i,S^*_{-i}) \leq (1+\meani-M_i) \frac{n}{\bk} + o(n)
\end{align}
\end{lemma}
\begin{proof}
We start with the following claim: 
\begin{claim}
Given $S^*_i$ and that the MAB algorithm is FATA and Monotonic, then for any possible strategy $S_i$ for an arm $i \in \armstop$, we have that $\E[\totpulls{i}(n)]\leq \frac{n}{\bk}+o(n)$. 
\end{claim}
\begin{proof}
Since the algorithm is monotonic, it follows that arm $i$ can maximize its number of pulls by always giving a reward of $\maxall$ whenever pulled. Further, since it is FATA and the other arms in $\armstop$ follow $S^*_i$, then $\E[\totpulls{i}(n)]\leq \frac{n}{\bk}+o(n)$. 
\end{proof}
For arm $i$ to have a utility $u_i=\Omega(n)$, then $\E[T_i(n)] \ge \alpha n$ where $\alpha$ is a constant. Since the other arms in $\armsup$ follow strategy $S^*_{-i}$, then we can reduce this instance to a MAB instance where the maximum honest agent has a mean of $\honestmean=\maxall$ and $\honestgapi=\maxall-\meani$, thus by the sharp adaptivity property it follows that if $\E[T_i(n)] \ge \alpha n$, then $\E[C_i(n)] \ge \alpha (\maxall-\meani)n - o(n)$ and therefore $u_i(S_i,S^*_{-i})\leq \alpha (1+\meani-\maxall) n + o(n)$. By the above lemma $\alpha$ has a maximum value of $\alpha=\frac{1}{\bk}$ and therefore  $u_i(S_i,S^*_{-i})\leq  (1+\meani-\maxall) \frac{n}{\bk} + o(n)$.    
\end{proof}
Finally, from the lower and upper bounds for the utilities of arms $i \in \armstop$ under $S^*$ holds regardless of the strategies of arms $i' \in \arms-\armsup$. Therefore, setting $\armstop=\armseq$ and $\arms-\armstop=\armsnon$. We have a partial approximate equilibrium as in Definition \ref{def:partial_approx_equilib}. 
\end{proof}

\thspucb*
\begin{proof}
We consider three cases, $M_i<m'$, $M_i=m'$ and $M_i>m'$ separately.

\paragraph{\textbf{Case ($M_i < m'$):}} 
Consider the case where arm $i$ has $M_i < m'$. Each arm $j \neq i$ follows the equilibrium strategy $\spucb$. Then, there exists at least one arm $j$ such that $M_j \ge m'$. It follows that we have a MAB instance where the optimal arm has a mean of $\mu^*=\maxall \ge m_i>M_i$. Therefore, since we are using a SAMF algorithm by monotinicity it follows that giving a reward of $M_i$ will maximize $\E[T^{SAMF}_i(n)]$ where $T^{SAMF}_i(n)$ is the number of pulls $i$ receives in the SAMF phase, further by the upper bounds assumed for the SAMF algorithm we have $\E[T^{SAMF}_i(n)]\leq \frac{c \ln(n)}{(\maxall-M_i)^2}$. Therefore, if $i$ follows the equilibrium strategy, then we have 
\begin{align*}
    u^{\text{SP+SAMF}}_i & = (1+\meani-M_i) + O\big(\ln(n)\big)+  
    \big(1+\indic[i \notin \armshonest] \cdot \mu_i \big) M_i \big(\ln(n)\big)^{\rho+3} \\
    & = \big(1+\indic[i \notin \armshonest] \cdot \mu_i \big) M_i \big(\ln(n)\big)^{\rho+3} 
    + O\big(\ln(n)\big).
\end{align*}
% $u^{\text{SP+SAMF}}_i= (1+\meani-M_i) + O\big(\ln(n)\big)+ \big(1+\indic[i \notin \armshonest] \cdot \mu_i \big) M_i \big(\ln(n)\big)^{\rho+3} = \big(1+\indic[i \notin \armshonest] \cdot \mu_i \big) M_i \big(\ln(n)\big)^{\rho+3} + O\big(\ln(n)\big)$. 

Now we consider arm $i$ deviating from the equilibrium strategy to $S'_i$. If arm $i$ bids $m'_i$ and if it gives an average of reward of $\hmeani'$ in the SAMF phase, then it follows that  
\begin{align*}
    u'_i\leq (1+\meani-m_i) + O\big(\ln(n)\big)  
    + \big(1+\indic[i \notin \armshonest] \cdot \mu_i \big) 
    \hmeani' \big(\ln(n)\big)^{\rho+3}.
\end{align*}
% $u'_i\leq (1+\meani-m_i) + O\big(\ln(n)\big) + \big(1+\indic[i \notin \armshonest] \cdot \mu_i \big) \hmeani' \big(\ln(n)\big)^{\rho+3}$. 
Since $\hmeani' \leq M_i$, the if follows that $\lim_{n \rightarrow \infty} \frac{u'_i}{u_i} \leq 1$ which means that the $S^{\text{SP+SAMF}}_i$ strategy is an equilibrium.

\paragraph{\textbf{Case ($M_i = m'$):}} 
In this case, we separate \emph{two possibilities}: (1) there exists an arm $j$ such that $M_j >m'$ and (2) $M_j =m'$.
First, suppose that $M_j > m'$. Since there exists an arm $j$ such that $M_j >m'$, under strategy profile $\spucb$ arm $j$ will give rewards of $m'+\frac{1}{\ln(n)}$. Therefore, the expected number of pulls in the SAMF phase can be upper bounded by $\E[T^{SAMF}_i(n)]\leq \frac{c \ln(n)}{(\ln(n))^2} = O\big((\ln(n))^3\big)$. Following similar analysis to the previous case we find that 
\begin{align*}
    u^{\text{SP+SAMF}}_i
    &= 
    (1+\meani-M_i) + O\big((\ln(n))^3\big) + \big(1+\indic[i \notin \armshonest] \cdot \mu_i \big) M_i \big(\ln(n)\big)^{\rho+3}
    \\
    &= \big(1+\indic[i \notin \armshonest] \cdot \mu_i \big) M_i \big(\ln(n)\big)^{\rho+3} + O\big((\ln(n))^3\big)
\end{align*}
% $u^{\text{SP+SAMF}}_i= (1+\meani-M_i) + O\big((\ln(n))^3\big) + \big(1+\indic[i \notin \armshonest] \cdot \mu_i \big) M_i \big(\ln(n)\big)^{\rho+3}= \big(1+\indic[i \notin \armshonest] \cdot \mu_i \big) M_i \big(\ln(n)\big)^{\rho+3} + O\big((\ln(n))^3\big)$. 
Similar to the previous case, if $i$ was to deviate to $S'_i$, then we would have 
\begin{align*}
    u'_i= \big(1+\indic[i \notin \armshonest] \cdot \mu_i \big) \hmeani' \big(\ln(n)\big)^{\rho+3} + O\big((\ln(n))^3\big).
\end{align*}
% $u'_i= \big(1+\indic[i \notin \armshonest] \cdot \mu_i \big) \hmeani' \big(\ln(n)\big)^{\rho+3} + O\big((\ln(n))^3\big)$ 
Therefore, we always have $\lim_{n \rightarrow \infty} \frac{u'_i}{u_i} \leq 1$ and therefore $S^{\text{SP+SAMF}}_i$ is an equilibrium.

Now consider the second case where there exists another arm $j \neq i$ with $M_j =m'$. If arm $i$ follows the equilibrium strategy of always giving $M_i$ in the SAMF phase, then by the FATA property of SAMF algorithms we have $\E[T^{\text{SAMF}}_i(n)]\ge\frac{n}{\bk}-o(n)$. Following similar analysis we have 
\begin{align*}
    u^{\text{SP+SAMF}}_i= (1+\meani-M_i) +(1+\meani-M_i) \frac{n}{\bk} -o(n) + \big(1+\indic[i \notin \armshonest] \cdot \mu_i \big) M_i \big(\ln(n)\big)^{\rho+3} 
\end{align*}
% $u^{\text{SP+SAMF}}_i= (1+\meani-M_i) +(1+\meani-M_i) \frac{n}{\bk} -o(n) + \big(1+\indic[i \notin \armshonest] \cdot \mu_i \big) M_i \big(\ln(n)\big)^{\rho+3}$. 
Therefore, $u_i \ge (1+\meani-M_i) \frac{n}{\bk} -o(n)$. 
Now we consider arm $i$ deviating to $S'_i$, then similar to Lemma \ref{lemma:pi_util_upper_bound} since we have a SAMF MAB algorithm the highest utility that arm $i$ can get is $u'_i\leq (1+\meani-M_i)\frac{n}{\bk} + o(n)$. Therefore, we always have $\lim_{n \rightarrow \infty} \frac{u'_i}{u_i} \leq 1$ and therefore $S^{\text{SP+SAMF}}_i$ is an equilibrium.

\paragraph{\textbf{Case ($M_i > m'$):}} Under the strategy profile, this means that arm $i$ is the only arm with $M_i=\maxall$ and therefore  $\bk=1$. If arm $i$ follows the equilibrium strategy, then it will be the unique optimal arm in the SAMF phase and therefore it would be pulled for an expected $n-o(n)$ rounds and its utility would be $u_i \ge (1+\meani-m') n-o(n)$. Suppose $i$ deviates to another strategy $S'_i$. Either $m'$ will change or it would not. Suppose $m'$ remains the same, then there must be an arm $j \neq i$ that always gives a reward of $m'$ and therefore by the sharp adaptivity property of SAMF algorithms $u'_i\leq (1+\meani-m') n-o(n)$ and therefore $\lim_{n \rightarrow \infty} \frac{u'_i}{u_i} = 1$. If $m'$ changes, then a collection of facts are implied (see Figure \ref{fig:SPUCB} for an illustration): (1) The new value of $m'$ which we will call $m'_{\text{new}}$ is smaller than $m'$ (i.e., $m'_{\text{new}}<m'$). (2) It must be the case that arm $i$ did not bid truthfully and gave a value $m_i\leq m'_{\text{new}}<m'$.(3) There must be another arm $j \neq i$ for which $M_j >m'_{\text{new}}$. Since $j$ would follow the equilibrium strategy, it would give rewards of $m'_{\text{new}}+\frac{1}{\ln(n)}$. If arm $i$ gives a reward greater than $m'_{\text{new}}$ during the SAMF phase, then it would be blocked (line 4A). Therefore, since arm $j$ always gives rewards of $m'_{\text{new}}+\frac{1}{\ln(n)}$, we obtain
\begin{align*}
    \E[\totpulls{i}(n)]=O((\ln{n})^{\rho+3})+\frac{2 \ln(n)}{(\ln(n))^2}=O((\ln{n})^{\rho+3})=o(n).
\end{align*}
% $\E[\totpulls{i}(n)]=O((\ln{n})^{\rho+3})+\frac{2 \ln(n)}{(\ln(n))^2}=O((\ln{n})^{\rho+3})=o(n)$ 
Hence, we have $u'_i=o(n)$ and deviation does not increase utility here.
\begin{figure}
  \centering
  \includegraphics[scale=0.5]{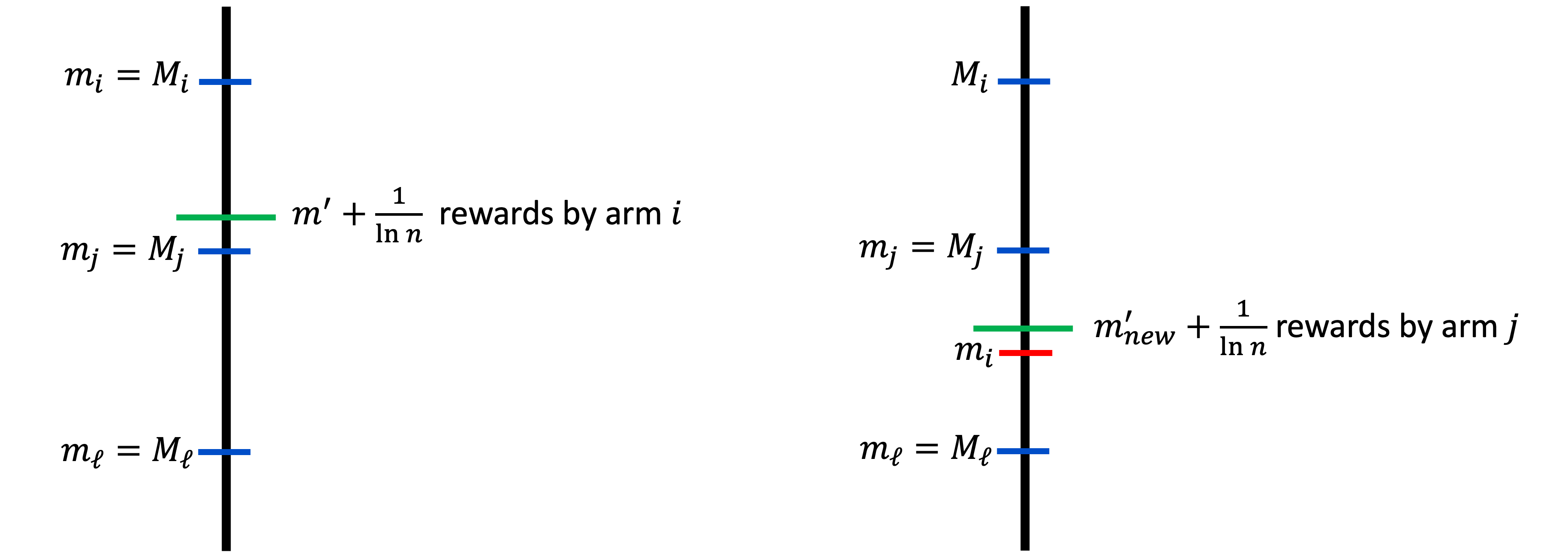}
  \caption{We have arms $i,j$ and $\ell$. On the left, we have truthful bidding resulting in arm $i$ giving rewards of $m'+\frac{1}{\ln(n)}$. On the right, when have arm $i$ under-bidding which causes $m'$ to change to $m'_{\text{new}}$ and for arm $j$ to instead give rewards at $m'_{\text{new}}+\frac{1}{\ln(n)}$.}
  \label{fig:SPUCB}
\end{figure}
\end{proof}

\sprpiguaren*
\begin{proof}
We use the equilibrium strategy $S^{SP+SAMF}$ of Theorem \ref{thm:sp_ucb}. The proof of point (1) follows immedialty. For point (2), note that an honest agent always bids truthfully therefore she would never be blocked. Further, since we use a SAMF algorithm then by sharp adaptivity $P(n)\ge \honestmean n-o(n)$ for any strategy profile (Theorem \ref{th:sr_always_sb_regret}).
\end{proof}

We restate the next theorem and give its proof
\thpuresp*
\begin{proof}
We consider an instance where all of the agents are honest, i.e., $\armshonest=\arms$. Further, all agents have the same mean value ($\mu_i=\honestmean$) and all of the agents have the same max performance value ($M_i=\maxall$). We further assume that for all agents purely sampling the reward without modification will never lead to $r_i=M_i$, i.e., $\forall i \in \arms: \Pr\limits_{r_i \sim \Dist_i} (r_i=M_i)=0$. 

Since we assume that the bidding is done truthfully and that $M_i=\maxall$, it follows that all arms will be included in the set $\armsmp$ ($\armsmp=\arms$) and that $m'=\maxall=M_i$. The algorithm pulls arms from $\armsmp$ and deletes an arm from $\armsmp$ only if it gives a value less than $\maxall=M_i$. Clearly, under an equilibrium strategy profile each arm gives a value of $M_i$ when pulled in the rounds of line 5 (see Appendix \ref{app:pure_sp} for more details). We begin with the following claim which lower bounds the number of pulls an arm receives until it is deleted. 
\begin{claim}\label{cl:bound_pulls_per_arm}
Set $\tau = \ceil{\frac{2 \ln{(n)}}{\ln(\frac{1}{1-\epsilon})}}$ then $\Pr(\text{arm $i$ is pulled at most $\tau$ times before it is deleted}) \ge 1 - \frac{1}{n^2}$.
\end{claim}
\begin{proof}
\begin{align*}
    \Pr(\text{arm $i$ is pulled at most $\tau$ times before it is deleted}) & = \Pr(\text{arm $i$ is deleted after 1 pull}) + \Pr(\text{arm $2$ is deleted after 1 pull}) \\ 
    & + \dots + \Pr(\text{arm $i$ is deleted after $\tau$ pull}) \\ 
    & = \epsilon + (1-\epsilon) \cdot \epsilon + (1-\epsilon)^2 \cdot \epsilon + \dots + (1-\epsilon)^{\tau-1} \cdot \epsilon \\
    & = \epsilon \frac{1-(1-\epsilon)^\tau}{1-(1-\epsilon)} \\ 
    & = 1-(1-\epsilon)^\tau \\ 
    & \ge  1-(1-\epsilon)^{\frac{2 \ln(n)}{\ln(\frac{1}{1-\epsilon})}} \\ 
    & = 1 - \frac{1}{n^2} 
\end{align*}
\end{proof}
By the above claim, it is clear that all arms will be deleted with high probability after each receives at most $\tau$ many pulls. Specifically, we have
\begin{align*}
    \Pr(\text{all arms are deleted after each receives at most$\tau$ many pulls}) & \ge (1-\frac{1}{n^2})^k \quad \quad \text{(by Claim \ref{cl:bound_pulls_per_arm})} \\ 
    & \ge 1 - \frac{k}{n^2} \quad \quad \text{(by Bernoulli's inequality)} \\ 
    & \ge 1 - \frac{1}{n}
\end{align*}
By the above with probability at least $\frac{1}{n}$ the revenue is at most $P(n) \leq k \cdot \tau + 2k = k \frac{2 \ln{(n)}}{\ln(\frac{1}{1-\epsilon})} + 2k = O(k \ln (n))$.

\end{proof}
\section{Further Remark on The Blocking Condition Line (4A) in SP+SAMF (Algorithm \ref{alg:spa_UCB_mech})}\label{app:blocking_cond}
\paragraph{Why do we need to modify the SAMF algorithm by blocking arm $i$ if $m_i \leq m'$ and $r_i> m'$?} To illustrate the reason, consider the following example where we have three arms $k=3$ all having the same mean of $0.1$ ($\mu=\meani=0.1 \ \forall i\in [k]$) and let the maximum reward values be the following: $M_1=1$, $M_2=0.8$, $M_3=0.3$. Suppose, we use the same mechanism as SP+SAMF (\ref{alg:spa_UCB_mech}) but with the blocking condition (line 4A) removed. Now, if we consider the same strategy profile $\spucb$, then arms 2 and 3 would bid $m_2=0.8$ and $m_3=0.3$, suppose arm 1 deviates and bids untruthfully with value $0.1$, then $m'=0.3$. It follows that arm 2 would give rewards of $0.3+\frac{1}{\ln{n}}$, but arm 1 can now give rewards greater than $0.3+\frac{1}{\ln{n}}$ such as $0.4$ and gain a linear utility of $u'_i=n-o(n) (1-(0.4-\mu_i))$. On the other hand, if arm 1 bids truthfully it can gain at most $u_i=n-o(n) (1-(0.8-\mu_i))$ since we would instead have $m'=0.8$. Therefore, clearly $\lim\limits_{n \xrightarrow[]{} \infty} u'_i/u_i >1$. However, with the blocking condition (line 4A) that cannot happen since arm 1 would be blocked. 

\section{SP+SAMF: $\bk=1$ and arm $i$ with $M_i=\maxall$ is an honest agent}\label{app:unique_top_is_honest}
Suppose that $\bk=1$ and the arm  $i$ with $M_i=\maxall$ is an honest agent, then the issue is that the equilibrium strategy $S_i^{\textsc{SP+SAMF}}$ would be to give in every SAMF MAB round $S_i^{\textsc{SP+SAMF}}=m'+\frac{1}{\ln(n)}$ but since $\maxall = M_i > m'+\frac{1}{\ln(n)}$ and $Support(\Dist_i) \subset [0,\maxi]$ this could mean that $i$ would have to absorb the reward in some rounds. Therefore, we introduce a special strategy for this case where the reward is never absorbed. 

Clearly, we assume that $\honestmean < m'+\frac{1}{\ln(n)}$ as otherwise trivially agent $i$'s optimal strategy is to never give any effort $c_i(t)=0$ whenever pulled. For convenience, let $\barm = m'+ \frac{1}{\ln(n)}$. We represent $\Dist_i$ as a mixture distribution where with probability $\probdown$ we sample $r< \barm $ from $\Distdown$ with its expectation denoted by $\meandown$ and with probability $\probup=1-\probdown$ we sample $r \ge \barm $ from $\Distup$ with its expectation denoted by $\meanup$. Accordingly, $\honestmean = \probup \meanup + \probdown \meandown$. The strategy is the following: 

\paragraph{Case $\probup \ge \frac{\barm - \meandown}{M_i}$:}  If $r \sim \Distup$, then with probability $p$ modify it to $M_i$ otherwise do not change the reward where $p=\frac{\barm-\probdown \meandown- \probup \meanup}{\probup (M_i-\meanup)}$. 

\paragraph{Case $\probup < \frac{\barm - \meandown}{M_i}$:} If $r \sim \Distup$, then always modify it to $M_i$. Further, if $r \sim \Distdown$ then modify it to $\barm$ with probability $p$ where $p=\frac{\barm - \probup M_i - \probdown \meandown}{\probdown (\barm-\meandown)}$. 

Note that both strategies never degrade the rewards. Further, since sampling in each round is independent of the history and uses the same distribution $\Dist_i$ it is i.i.d. Moreover, by simply plugging the value of $p$ according to the cases above, the new mean of the distribution is $\barm$. The rest of the proof for the utility follows identical steps.

\section{Equilibrium Guarantee of \textsc{Pure-SP} (Algorithm \ref{alg:pure_sp_mech})}\label{app:pure_sp}
Here we show that the \textsc{Pure-SP} mechanism leads to a revenue of $P(n) \ge M_{\text{top-1}} n - o(n)$ under an equilibrium. We start by describing the equilibrium strategy profile. For simplicity, we assume that if $\bk=1$ then the only agent with $M_i=\maxall$ is not $\honestarm$.

% Below we give a full description of the equilibrium strategy profile. For convenience, we assume below that if $\bk=1$, then the only agent with $M_i=\maxall$ is not $\honestarm$ (the honest agent with the maximum mean), this case leads to identical guarantees but using a more complicated strategy for the honest agent $\honestarm$ since we assume that honest agents would never absorb (degrade) their rewards (see Appendix \ref{app:unique_top_is_honest}).

\begin{restatable}{theorem}{thpure_mech}\label{thm:pure_sp_mech}
Under \textsc{Pure-SP} (Algorithm \ref{alg:pure_sp_mech}) strategy profile $\spure$ is an asymptotic equilibrium that consists of the following: (1) In the bidding phase (line 1) each arm bids truthfully with $m_i=M_i$, (2) In the reward phase (line 6), all rewards are absorbed if the agent is not honest otherwise no positive effort is added. (3) In (line 5) phase each arm gives a reward of $m'$ if pulled.   
\end{restatable}
\begin{proof}
\textbf{Case ($i \notin \armsmp$):} \ Under this case the maximum achievable utility for arm $i$ if it is not honest will be 
\begin{align*}
    u_i(m_i) = 1-(m_i-\mu_i) + 2 m_i = 1+\mu_i + m_i 
\end{align*}
Since the $u_i(m_i)$ is increasing in the bid value $m_i$ then clearly the agent should bid truthfully. If the agent is honest then he will always bid truthfully further, in the reward phase of (line 5) his utility is clearly maximized by not giving any positive effort. 

\paragraph{\textbf{Case ($i \in \armsmp$):}} Since we care about an asymptotic equilibrium we will ignore the rounds of lines (1 and 6) for the bidding and reward phases since they are constant whereas the pulls of line (5) are linear $\Omega(n)$. If the agent follows the strategy $\spure$ of always giving $M_i$ in line (5) then the utility will be
\begin{align*}
    u^\textsc{Pure-SP}_i = \ell \cdot (1-(M_i-\mu_i)) = \ell \cdot (1+\mu_i-M_i) 
\end{align*}
where $\ell_i$ is the number of pulls in that phase. Clearly, $\floor{\frac{n-3k}{|\armsmp|}} \leq \ell_i \leq \ceil{\frac{n-3k}{|\armsmp|}}$ since the algorithm cycles among the arms in $\armsmp$. The deviation strategy would give no effort in the last pull $\ell_i$ leading to a maximum utility of at most $u_i'$. Since $u_i' \leq u^\textsc{Pure-SP}_i+ 1$ it follows that strategy $S^\textsc{Pure-SP}$ leads to an asymptotic equilibrium. 
\end{proof}

Note that under the above equilibrium $\armsmp=\armstop$ and since it is an equilibrium for these arms to give a reward of $M_{\text{top-1}}$ in the phase of line (5) in \textsc{Pure-SP} it follows that the revenue is $P(n) \ge M_{\text{top-1}} n - o(n)$.

\section{Results for General Cost Functions}\label{sec:cost}
Up until now for any given round $t$ the utility of the arm for this round is $u_i(t)=0$ if the arm is not pulled or $u_i(t)=1-c_i(t)\ge 0$ where $c_i(t)$ is the effort positive or negative. Implicitly the cost is exactly equal to the effort, but one can see that in general the relation between the effort and the cost could be more complicated. Specifically, if an arm $i$ improves over the reward by $c_i(t)$, then the cost it incurs in round $t$ is $f_i(c_i(t))$ where $f_i(.)$ is a \emph{cost function} specific to agent $i$. Accordingly, the utility becomes: 
\begin{align}
    u_i = \E[\totpulls{i}(n)] - \sum_{t=1}^n \E[f_i(c_i(t))]
\end{align}

% Note that this implies that the total utility (for the whole horizon) of the arm is non-negative. Therefore, \emph{individual rationality} is trivially satisfied, i.e. $u_i(S)\ge 0$ where $S$ is any strategy profile. However, introducing cost functions makes individual rationality not trivial to satisfy. Specifically, if an arm $i$ improves over the reward by $c_i(t)$, then the cost it incurs in round $t$ is $f_i(c_i(t))$. 
It follows that in the original cost (utility) model we had $\forall i \in [k]:f_i(c_i(t))=c_i(t)$. Clearly, depending on the function $f_i(.)$ and the effort, the arm might incur negative utility. We assume two reasonable properties for $f_i(.)$: (1) $f_i(0)=0$. and (2) $f_i$ is increasing. We further define $g_i(x)=\E_{r\sim \Dist_i}[f_i(x-r)]$ where $\Dist_i$ is the reward distribution of arm $i$. It follows that $g_i$ has a domain of $dom(g_i)=\{ x| M_i \ge x\ge 0\}$. Further, we assume that $f_i$ and $g_i$ have no dependence on the horizon. Since $f_i$ is increasing then $g_i$ is also increasing. For mathematical convenience, we assume the following condition on the realized rewards and modified rewards which is that they lie in a finite discrete space that is independent of the horizon:
\begin{condition}\label{cond:discrete_vals}
The realized and modified rewards belong to a set $\Omega$ which is finite and discrete and $\Omega \subset [0,1]$. Further, its size is independent of the horizon $|\Omega|=\Theta(1)$. 
\end{condition}
The size of the set can be large but independent of the horizon. We note that such an assumption is simple and holds in real life. For example, a movie rating website could give a rating from 1 to 10 leading to $|\Omega|=10$.
We now define $\mic$ which is the maximum performance level at which the expected utility remains non-negative:
\begin{align}
    \mic = \max_{x \le M_i} \{x:1-g_i(x)>0\}.
\end{align}
Note in the above that without Condition (\ref{cond:discrete_vals}) we would have to use a supremum and end up with more cumbersome and less clear descriptions. We now define a \emph{sustainable arm} and an \emph{unsustainable arm}:
\begin{definition}
\textbf{Sustainable and Unsustainable Arms:} We call an arm $i$ sustainable if $\mic=M_i$ and unsustainable if $\mic<M_i$. 
\end{definition}
In words, sustainable arm can sustain performing at their maximum level since they gain positive utility in expectation but that is not the case for unsustainable arms. Accordingly, we define the sustainable setting where arms cannot incur negative utility since $M_i=M^f_i$. 
\begin{definition}\label{def:Sustainable_Setting}
\textbf{Sustainable Setting:} In the sustainable setting for each arm $i \in \arms: M_i=M^f_i$. 
\end{definition}

% We also define the following $\maxallc=\max_{i\in [k]} \mic$ and $\armstopc  =\{i \in [k]: \mic = \maxallc \}$ and $\kic=|\armstopc|$. We also define $S^{*}_i$ as the strategy of always giving a reward of $\mic$
\subsection{Cost Functions in the $\CompInfoInitCap$ and Sustainable Setting: Non-Existence of Sub-Optimal Revenue Equilibria}
In this subsection we are in the sustainable setting and therefore $\forall i \in \arms: M_i=M^f_i$. While this implies that any arm will always have non-negative utility, this setting allows for a large set of possible cost functions. We also assume that agents are $\compInfo$. We now show similar to Subsection (\ref{subsec:no_bad_equilb}) that there does not exist an asymptotic equilibrium where the principal obtains revenue below $\maxallc$ for MAB algorithms that are sharply adaptive and monotonic. We need a condition similar to Condition \ref{cond:cond_good_revenue}:  
% \begin{condition}\label{cond:cost_cond_good_revenue}
% $\min_{i \in \armstopc} \Big(1-g_i(\maxallc)\Big) > \frac{1}{\bk}$. 
% \end{condition}
\begin{condition}\label{cond:cost_cond_good_revenue}
$\bk > (1+\epsilon) \min\limits_{i \in \armstopc} \frac{1-g_i(0)}{1-g_i(\maxallc)} $ where $\epsilon>0$ is a constant. 
\end{condition}

% \begin{theorem}\label{th:sr_np_bad_equilib}
% Given a MAB algorithm that satisfies sharp adaptivity and monotonicity, if Condition (\ref{cond:cond_good_revenue}) is satisfied, then there exists no asymptotic equilibrium strategy profile $S$ such that $P(n)\leq \alpha \maxall n +o(n)$ where $\alpha <1$. 
% \end{theorem}

\begin{theorem}\label{th:cost_sr_np_bad_equilib}
Given a MAB algorithm that satisfies sharp adaptivity and monotonicity, if Condition (\ref{cond:cost_cond_good_revenue}) is satisfied, then there exists no partial asymptotic equilibrium strategy profile $S$ such that $P(n,S)\leq \alpha \maxallc n +o(n)$ where $\alpha <1$. 
\end{theorem}
\begin{proof}
The proof follows by identical steps to Theorem \ref{th:sr_np_bad_equilib}. Except that $i$'s utility under strategy profile $S$ would be upper bounded by:
\begin{align*}
        u_i(S) \leq \toptj{i}(S) (1-g_i(\maxall)) + \frac{\tsubopt(S)}{\bk} (1-g_i(0)) + o(n) 
\end{align*}
Further, under the deviation strategy profile $S'=(S^*_i,S_{-i})$, we get the following lower bound for $i$'s utility: 
\begin{align*}
        u_i(S') \ge \toptj{i}(S) (1-g_i(\maxall)) + \tsubopt(S) (1-g_i(\maxall)) -o(n)
\end{align*}
Now, Condition \ref{cond:cost_cond_good_revenue} states that $\bk > (1+\epsilon) \min\limits_{i \in \armstopc} \frac{1-g_i(0)}{1-g_i(\maxallc)} $, then we have $1-g_i(\maxallc) > (1+\epsilon)  \frac{1-g_i(0)}{\bk} > \frac{1-g_i(0)}{\bk} + \Theta(1)$, since the original utility is $u_i(S) \leq \toptj{i}(S) (1-g_i(\maxall)) + \frac{\tsubopt(S)}{\bk} (1-g_i(0)) + o(n)$. It follows that $\lim\limits_{n \xrightarrow[]{}\infty} \frac{u_i(S')}{u_i(S)}=\infty$. Therefore, this is not a partial asymptotic equilibrium.
\end{proof}

\subsection{Linear Cost Functions: Achieving Top Performance Equilibrium for any Value of $\bk$}
Here we consider a simple but meaningful cost function which is the linear cost function. Specifically, $\forall i \in [k]: f_i(x)=a_i x$ where $a_i>0$ is a coefficient that is known to the agent but not necessarily to the principal. Clearly, in the plain cost model, we had $a_i=1$ for all arms. Accordingly, $g_i(x)=a_i(x-\meani)$ and therefore $\mic$ is: 
\begin{align}
    \mic = \max_{x\in \Omega, x \le M_i} \{x:1-a_i (x-\meani)>0\}.
\end{align}

We show that in the Sustainable Setting when agents are $\compInfo$, then a SAMF algorithm can also achieve a top performance equilibrium regardless of whether Condition \ref{cond:cost_cond_good_revenue} holds or not (Theorem \ref{thm:cost_info}). We end however by showing that if there exists $i: M_i>M^f_i$ then such an equilibrium does not necessarily exist (Theorem \ref{thm:neg_cost}).

\begin{theorem}\label{thm:cost_info}
In the $\compInfo$ and sustainable setting where $\forall i\in[k]: M_i=M^f_i$, if we have a SAMF MAB algorithm, then the strategy profile $S^*$ where $\forall i \in \armsup: S_i=S^*_i$ is a partial asymptotic equilibrium leading to optimal revenue $P(n,S^*)=\maxall n -o(n)$.
% exists a partial asymptotic equilibrium where $\armseq=\armstop$, $\armsnon=\arms-\armstop$ and $\Seq = S^*_{\text{top}}$ where $\forall i \in \armstopc: S_i=S^{*}_i$. Further, this partial equilibrium leads to optimal revenue with $P(n,\Seq,\Snon)=\maxall n -o(n)$. 
\end{theorem}

% Given the $\compInfo$ setting, if a MAB algorithm is SAMF, then there exists a strategy profile $S^*$ where $\forall i \in \armsup: S_i=S^*_i$ is a partial asymptotic equilibrium leading to optimal revenue $P(n,S^*)=\maxall n -o(n)$.
\begin{proof}
The proof is identical to the proof of Theorem \ref{th:pi_equilib}. Note however that the utility lower bound for an arm $i \in \armstop$ under $S^*=(S^{*}_i,S^{*}_{-i})$ is: 
\begin{align*}
    u_i(S^{*}_i,S^{*}_{-i}) \ge \Big(1-a_i(\maxallc-\meani)\Big) \frac{n}{\kic} - o(n) 
\end{align*}
Whereas under a deviation strategy profile $S'=(S_i,S^{*}_{-i})$ for an arm $i \in \armstop$ the utility upper bound is: 
\begin{align*}
        u_i(S_i,S^{*}_{-i}) \leq \Big(1-a_i(\maxallc-\meani)\Big) \frac{n}{\kic} + o(n) 
\end{align*}
Setting $\armseq=\armstop$ and  $\armsnon=\arms-\armstop$ we have a partial asymptotic equilibrium.   
\end{proof}

Now we show that if $\exists i \in \arms: M_i> M^f_i$ (we are not in the sustainable setting), then we may not necessarily have a top performance equilibrium even if we are using a SAMF algorithm.
\begin{theorem}\label{thm:neg_cost}
If $\exists i\in \arms: M_i> M^f_i$, then even when using SAMF MAB algorithm and dealing with the special case of the linear cost function, then strategy $S^{*}_i$ where each agent gives $M^f_i$ may not be an asymptotic equilibrium for the arms in $\armstopc$.  
\end{theorem}
\begin{proof}
Consider the $\epsilon$-Greedy algorithm where ties between top arms are broken by randomly sampling (as shown in algorithm (\ref{alg:eps_greedy})). Further, let $k=2$ and $\mu_1=\mu_2$ and $M^f_1=M^f_2$ but $M_1>M^f_1$ and $M_2=M^f_2$. We will show that profile $(S^{*}_1,S^{*}_2)$ is not an equilibrium. First, it is easy to see (by symmetry) that $u_1(S^{*}_1,S^{*}_2)=\frac{n}{2} \Big(1-a_1(\maxallc-\mu_1)\Big)$. 

Now suppose arm $1$ deviates to $S'_1$ where instead it gives $M_1>\maxall$ in the first pull and then always gives $\maxallc$. Then given that arm 2 always gives $\maxallc$, then it follows that $\hat{\mu}_{1}(t)>\hat{\mu}_{2}(t) \ \forall t\ge 2$ and therefore arm 1 will always be pulled for a linear number of rounds leading to $u_1(S'_1,S^{*}_2)= n \Big(1-a_1(\maxallc-\mu_1)\Big)-o(n)$. Therefore,  $(S^{*}_1,S^{*}_2)$ is not an equilibrium. 
\end{proof}

\section{Useful Theorems}
\begin{theorem}[Hoeffding's Inequality]\label{th:hoffbound}
Suppose $X_1,\dots,X_n$ is a collection $n$ many $[0,1]$ i.i.d. sampled values from a distribution $\Dist$ of expected value $\mu$. Let $\hat{\mu}=\frac{\sum_{i=1}^n X_i}{n}$, then we have the following: 
\begin{align}
    \Pr(|\hat{\mu}-\mu| \ge \epsilon) \leq 2 \exp(-2 n \epsilon^2)
\end{align}
\end{theorem}

\begin{theorem}[Chernoff Bound]\label{th:cherbound}
Let $X_1,\dots,X_n$ be independent $0$-$1$ random variables with $S_n=\sum_{i=1}^n X_i$ , then for $0< \delta \leq 1$ we have: 
\begin{align}
    \Pr(S_n \leq (1-\delta) \E[S_n]) \leq \exp(-\frac{\delta^2  \E[S_n]}{2})
\end{align}
\end{theorem}

\end{document}